\documentclass[aps,prb,reprint,footinbib,floatfix]{revtex4-1}
\usepackage{amsmath}
\usepackage{amssymb}
\usepackage{graphicx}

\usepackage[usenames]{color}
\usepackage[normalem]{ulem}

\newcommand{\VEC}[1]{\mathbf{#1}}
\newcommand{\eq} [1]{Eq.~(\ref{#1})}
\newcommand{\bra}[1]{\left\langle\,#1\,\right|}
\newcommand{\ket}[1]{\left|\,#1\,\right\rangle}

\newcommand{\KK}{\mathbf{k}_\parallel}
\newcommand{\kk}{k_\parallel}

\begin{document}

\title{Microscopic origin of the relativistic splitting of surface states}

\author{E. E. Krasovskii} 
\affiliation{Departamento de F\'{i}sica de Materiales, Facultad de
  Ciencias Qu\'{i}imicas, Universidad del Pais Vasco/Euskal Herriko
  Unibertsitatea, San Sebasti\'{a}n/Donostia, Spain}
\affiliation{Donostia International Physics Center (DIPC),
  San Sebasti\'{a}n/Donostia, Spain}
\affiliation{IKERBASQUE, Basque Foundation for Science, Bilbao, Spain}

\begin{abstract}
Spin-orbit splitting of surface states is analyzed within and beyond 
the Rashba model using  as examples the (111) surfaces of noble metals, 
Ag$_2$Bi surface alloy, and topological insulator Bi$_2$Se$_3$. The 
{\it ab initio} analysis of relativistic velocity proves the Rashba model 
to be fundamentally inapplicable to real crystals. The splitting is found 
to be primarily due to a spin-orbit induced in-plane modification of the 
wave function, namely to its effect on the {\em nonrelativistic} Hamiltonian. 
The usual Rashba splitting -- given by charge distribution asymmetry 
-- is an order of magnitude smaller.  
  \end{abstract}

\maketitle

\section{INTRODUCTION}
Spin structure of nonmagnetic surfaces has attracted much interest in the last 
decade due to the promising properties of spin-split surface states for spintronics 
applications.~\cite{Wolf_science2001,zutic2004} The idea of using the Rashba 
effect~\cite{RB1984JL,RB1984JPh} in the Datta-Das spin transistor~\cite{DD1990} 
and for spin filtering~\cite{Koga2002} has stimulated great activity in the 
search of enhanced spin-orbit splitting at surfaces. Upon the discovery in 1996 
of the splitting of the surfaces state on Au(111)~\cite{LaShell1996,Nicolay2001,Wissing2013} 
much attention has turned to metal surfaces: spin-split surface states were 
promptly found on (110) surfaces of W and Mo covered with thin overlayers
\cite{Rotenberg1998,Rotenberg1999,Hochstrasser2002,Koitzsch2005,Schiller2005,
Shikin2008,Rybkin2012} and on clean surfaces of Bi and Sb.~\cite{Agergaard2001,
Koroteev2004,Sugawara2006,Kimura2010} The giant spin splitting found in Bi and 
Pb/Ag(111) surface alloys~\cite{Meier2008,Gierz2010,Ast2007,Ast2008} has inspired 
a search for a strong Rashba effect in heavy-element adsorbates on technologically more 
attractive (111) surfaces of Si and Ge.~\cite{Gierz2009,Sakamoto2009,Hatta2009,Yaji2010} 
Recently, a giant Rashba splitting was observed in ternary semiconductors 
BiTeI,~\cite{Ishizaka2011,Bahramy2011,Landolt2012,Crepaldi2012} 
BiTeCl,~\cite{Eremeev2012,Landolt2013} and BiTeBr.~\cite{Eremeev2013} 

\begin{figure*}[t]
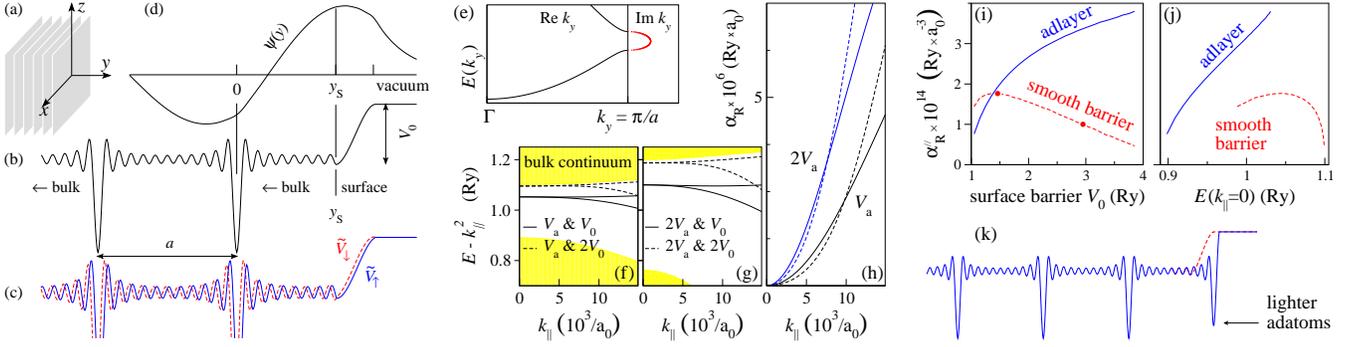
  
\includegraphics[height=0.19\textheight]{fig1aN.eps}\hspace{2mm}
\includegraphics[height=0.19\textheight]{fig1e.eps}\hspace{3mm}
\includegraphics[height=0.19\textheight,clip=true,trim=5 0 0 0]{fig1i.eps}
\caption{\label{model} (Color online) Microscopic Rashba model. 
(a)~Coordinate frame. (b)~Model potential $V(y)=-V_\text{a}\sum_{n}\cos(2\pi ny/a)$ 
with lattice constant $a=\pi a_0$, $V_\text{a}=0.1$~Ry, and vacuum level $V_0=1.5$~Ry. 
(c)~Effective potentials ${\tilde V}_{\downarrow}(y)={\tilde V}_{\uparrow}(-y)$ in
\eq{eq1d}. (d)~Surface state wave function. (e) Complex band structure: inside the 
gap it is $\text{Re}\,k_y=\pi/a$. (f)~Energy dispersion $E(\kk)$ of the surface 
states for the potential $V$ shown in graph (b). (g)~The same for the potential 
$2V$. 
(h)~$2\alpha_{\text R}=d(E_{\uparrow}-E_{\downarrow})/d\kk$ for the split surface 
states for the potential $V_0$ (solid lines) and 2$V_0$ (dashed lines). (i) Splitting 
parameter $\alpha_{\text R}''$ as a function of the surface barrier $V_0$ and (j) of the 
eigenenergy at $\bar\Gamma$  for the potential with and without a ``singularity'' 
at the surface, graph (k).}
\end{figure*}

The way to practical spintronics depends on a deep understanding of
fundamental mechanisms by which material properties determine the 
magnitude of the spin-orbit splitting. The early theoretical analyses of 
the splitting effect at the high-Z crystal surfaces~\cite{Petersen2000,
Henk2003,Hoesch2004,Bihlmayer2006} emphasized its similarity to the Rashba 
effect in semiconductor heterojunctions, which is well understood within 
the two-dimensional electron gas model.~\cite{RB1984JL,RB1984JPh} The model 
is described by the Rashba-Bychkov (RB) Hamiltonian:  
\begin{equation}
\label{rbh}
{\hat H}_\text{RB} = {\hat p_\parallel}^2/2m^* + \alpha_{\text R}\boldsymbol\sigma\cdot
({\mathbf n}\times\hat {\mathbf p}_\parallel).
\end{equation}
In application to semi-infinite crystals, the microscopic model behind 
Eq.~(\ref{rbh}) is obtained from the true two-component Hamiltonian
(in Rydberg atomic units)
\begin{equation}
\label{sreq}
\hat H = {\hat p}^2 + V(\VEC{r}) + \beta{\boldsymbol\sigma}\cdot 
\left [\,\boldsymbol\nabla V({\VEC r})\times\VEC{\hat p}\,\right ] 
\end{equation}
by neglecting the variation of the crystal potential $V({\VEC r})$ parallel 
to the surface, see Fig.~\ref{model}(a). Thus, equation~(\ref{rbh}) assumes a 
free motion along the surface, and the material dependence of the spin-orbit 
splitting comes through the Rashba parameter $\alpha_{\text R}$. In the RB 
model, $\alpha_{\text R}$ has a clear physical meaning of the expectation 
value of the potential gradient in the surface normal direction $\VEC{n}$ (along $y$ axis):
\begin{equation}
\label{rbp}
\alpha_{\text R} = \beta\bra{\psi}\partial V/\partial y\ket{\psi} =
\beta\int\limits_{-\infty}^{+\infty}\frac{\partial V}{\partial y}\,\rho(y)\;dy,
\end{equation}
with $\beta=\hbar/4m^2c^2$. Owing to the perfectly parabolic dispersion $E(\kk)$ of 
the Au(111) surface state~\cite{Nicolay2001} and to the nearly total spin polarization 
of the split branches, the Au(111) case is often considered  a textbook example of 
Rashba effect in metals and a demonstration of the applicability of Eq.~(\ref{rbh}). 
This view is detailed in Refs.~\onlinecite{Bihlmayer2006} and \onlinecite{Nagano2009}: 
since the spin-orbit interaction is significant only in a close vicinity of the nucleus, 
where the potential is spherically symmetric, the value of $\alpha_{\text R}$ is 
determined by the surface-perpendicular asymmetry of the probability density 
distribution $\rho(\VEC{r})=|\psi(\VEC{r})|^2$ at the nuclei. This idea, expressed 
by Eqs.~(\ref{rbh}) and (\ref{rbp}), has been the basis for the analysis of 
individual atomic contributions to the spin splitting and polarization at real 
surfaces.~\cite{Hatta2009,Yaji2010,Bentmann2011,Lee2012,Hortamani2012} In addition, 
in Refs.~\onlinecite{Premper2007,Ast2007,Ast2008,Gierz2009,Moreschini2009} an 
important role is attributed to the in-plane inversion asymmetry and to the related 
surface-parallel potential gradient. In spite of their different view on the 
relative importance of surface-normal and in-plane asymmetry, all the 
studies have focussed on the shape of $\rho(\VEC{r})$ and how it may affect 
the $\boldsymbol\nabla V$ expectation value, but so far no attempts have been 
made to express the asymmetry underlying the effect in energy units and to quantify 
different contributions to the spin-orbit splitting. The microscopic origin of the 
Rashba effect, i.e., its relation to the shape of the wave function in real crystals, 
thus remains an open question.

The aim of this work is to understand the physics behind the Rashba parameter 
in real crystals. An {\it ab initio} analysis of the Hamiltonian~(\ref{sreq}) 
is performed to reveal the actual value of the RB term. A surprising result 
is that the pure Rashba contribution to the splitting -- the one given by 
Eqs.~(\ref{rbh}) and (\ref{rbp}) -- typically yields only a few percent of 
the whole effect, while the major part arises from a relativistic modification 
of the wave function through its influence on the {\em nonrelativistic} energy 
operator.

\section{microscopic Rashba model}
Before we turn to real crystals let us take a closer look at how the 
microscopic Rashba model works. To be specific about terminology, here 
by Rashba model is understood a system in which electrons move freely 
along the surface. In the geometry of Fig.~\ref{model}(a), 
with the spin quantization axis along $z$ and with $\KK$ along $x$, 
the spatial and spin variables in the Schr\"odinger equation separate, 
and for a given $\kk$ the problem reduces to a pair of independent 
scalar equations for the two eigenfunctions $\psi_{\uparrow}(y)$ and 
$\psi_{\downarrow}(y)$:~\cite{Krasovskii2011}
\begin{equation}
\label{eq1d}
-\psi_{\uparrow\downarrow}''(y)+{\tilde V}_{\uparrow\downarrow}(y)\psi_{\uparrow\downarrow}(y) 
= (E-\kk^2)\psi_{\uparrow\downarrow}(y).
\end{equation}
The effective potential 
${\tilde V}_{\uparrow\downarrow}(y)=V(y)\pm\beta\kk V'_y(y)$ depends upon spin 
(``$-$'' for $V_\uparrow$ and ``$+$'' for $V_\downarrow$) and upon $\kk$. 
For a bulk crystal with inversion symmetry, $V(y)=V(-y)$, the effective 
potentials are related as ${\tilde V}_{\downarrow}(y)={\tilde V}_{\uparrow}(-y)$,
as illustrated by Figs.~\ref{model}(b) and \ref{model}(c) (bulk is to the 
left from the surface plane $y_{\text S}$). Equation~(\ref{eq1d}) 
then yields the same band structure $E(k_y)$ for both spins, for real 
and for complex $k_y$ [in the spectral gap, Fig.~\ref{model}(e)], which 
is known as Kramers degeneracy. The existence of a surface state in the 
gap depends on whether the regular solution of Eq.~(\ref{eq1d}) for 
$y>y_{\text S}$ matches the bulk evanescent wave at the same energy in 
both value and slope,~\cite{Tamm1932,Forstmann1970} see Fig.~\ref{model}(d). 
Being equivalent in an infinite crystal, the potentials ${\tilde V}_{\downarrow}$ 
and ${\tilde V}_{\uparrow}$ are different when looked at from the surface, 
see  Fig.~\ref{model}(c): ${\tilde V}_{\downarrow}(y)$ is slightly shifted to 
the left and ${\tilde V}_{\uparrow}(y)$ to the right relative to $V(y)$. 
Therefore, the evanescent waves of spin $\uparrow$ and spin $\downarrow$ 
are different at $y_{\text S}$, and the matching occurs at different energies 
for spin $\uparrow$ and spin $\downarrow$, which is known as Rashba splitting.

The dispersion lines $E(\kk)$ calculated in the nonperturbative 
complex-band-structure approach 
\cite{*[{The complex band structure is calculated by a variational plane 
wave method using the inverse band structure formalism described in }] 
[{. The solutions for $y>y_{\text S}$ are obtained with the Runge-Kutta 
method}] Krasovskii1997}
are presented in Fig.~\ref{model}(f) with full lines for the potential 
$V$ [shown in Fig.~\ref{model}(b)] and with dashed lines for the same 
bulk potential but for twice larger surface barrier, $V_0\to 2V_0$.
Figure~\ref{model}(g) shows the results for the potential $2V$ (i.e., both 
$V_0\to 2V_0$ and $V_\text{a}\to 2V_\text{a}$ for a larger ``atomic number''). 
The spin-orbit effect on the wave function leads to a dependence of 
$\alpha_{\text R}$ on $\kk$, see Fig.~\ref{model}(h), i.e., to a nonparabolicity 
of $E(\kk)$. [In the RB model~(\ref{rbh}) the slope at 
$\kk=0$ (point $\bar\Gamma$) equals zero,
\cite{[{This follows from the Ehrenfest theorem, 
as discussed in Chapter 6.2.2 in}] Winkler2003} so at finite $\kk$ we 
define 
$2\alpha_{\text R}=d(E_{\uparrow}-E_{\downarrow})/d\kk$.]
More important and counterintuitive result is that the potential 
scaling does not always increase the splitting -- it may even become 
smaller because at sufficiently small $\kk$ $\alpha_{\text R}$ decreases 
with increasing the barrier height, 
see Figs.~\ref{model}(f)--\ref{model}(h). Microscopically, this 
means that the effect of a larger potential gradient is completely cancelled by a 
modification of the wave function. From the complex band structure point of view 
this stems from the fact that in the $2V$ case [Fig.~\ref{model}(g)] the surface 
state is pushed toward the gap edge, where $\text{Im}\,k_y$ and, consequently, 
the wave function changes faster with energy, see Fig.~\ref{model}(e). 

The role of the surface is demonstrated by Fig.~\ref{model}(i), which
compares the splitting as a function 
of the surface barrier $V_0$
for a ``clean'' surface and for a surface with a ``lighter adatom'', see 
Fig.~\ref{model}(k). 
For small $\kk$ the splitting grows as $\alpha_{\text R}''\kk^3$, so  
Figs.~\ref{model}(i) and \ref{model}(j) show the coefficient $\alpha_{\text R}''$
determined by regression.
[The bulk potential is the one of Fig.~\ref{model}(b), 
so the two circles in the dashed curve in Fig.~\ref{model}(i) correspond to 
the data in Fig.~\ref{model}(f).] In both cases 
$\alpha_{\text R}''$ rapidly 
grows at small $V_0$, but at larger barriers it keeps growing in the 
adatom case and decreases for the smooth barrier.  
This seemingly different behavior, however, looks rather similar when 
$\alpha_{\text R}''$ is plotted as a function of the eigenenergy, see Fig.~\ref{model}(j): 
in both cases $\alpha_{\text R}''$ is seen to diminish when the energy is pushed 
to a gap edge, i.e., it shows the same trend as with scaling the potential, 
Figs.~\ref{model}(f) and \ref{model}(g). According to Eq.~(\ref{rbp}), the 
peculiar dependence of $\alpha_{\text R}$ on the system parameters results 
from a complicated redistribution of the charge density, whereby different
contributions to the potential gradient either cancel or enhance each other. 
A practical conclusion, however, can be formulated quite simply: in a Rashba 
system, to achieve larger splitting the surface state should be placed 
sufficiently far from the edges of the gap. Note, that here $\alpha_{\text R}$ 
is controlled through the modification of a {\em nonrelativistic} wave function: 
indeed, at $\kk=0$ it is ${\tilde V}_{\uparrow}={\tilde V}_{\downarrow}=V$, 
and the spin-orbit effect on the wave function vanishes.

\section{rashba splitting in real crystals}
\begin{figure}[t]  
\includegraphics[width=0.45\textwidth]{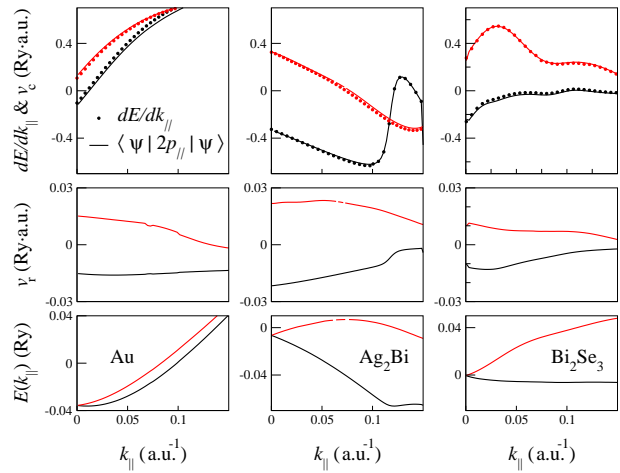}
\caption{\label{real} (Color online) Group velocity as a function of
the Bloch vector in the direction $\bar\Gamma\bar M$ for the surface 
states on Au(111), left; Bi/Ag(111), middle; and Bi$_2$Se$_3$, right 
column. In the upper row, circles show a numerical derivative of the 
eigenvalues $dE(\kk)/d\kk$, and solid lines show the nonrelativistic part 
$v_{\text c}$. The middle row shows the relativistic part $v_{\text r}$, 
and the bottom row the dispersion $E(\kk)$. Everywhere black color is 
used for the lower branch and red for the upper one. For Bi/Ag(111), 
the broken line for the upper branch at $\kk =0.07$~a.u.$^{-1}$ indicates 
the interpolated values at the place of avoided crossing with a higher 
lying surface state. 
}
\end{figure}

The fundamental difference between the spin-orbit splitting in real crystals
and in the RB model is best seen at the $\bar\Gamma$ point, where the
relativistic effect manifests itself as a nonzero group velocity $dE/d\kk$,
with the opposite sign for the two branches. The expression for $dE/d\kk$
follows from Eq.~(\ref{sreq}), cf. Eq.~(3) in Ref.~\onlinecite{Oguchi2009}:
\begin{equation}
\label{relvel}
dE/d\kk=2\bra{\psi}\hat p_\parallel\ket{\psi} + 
\beta\bra{\psi}{\boldsymbol\sigma}\cdot\left 
[\boldsymbol\nabla V\times\boldsymbol\tau_\parallel\right ]\ket{\psi},
\end{equation}
where $\boldsymbol\tau_\parallel={\VEC k}_\parallel/\kk$ and 
$\hat p_\parallel = \hat{\VEC p}\cdot\boldsymbol\tau_\parallel$. In 
the RB model, the wave function is 
$\psi({\VEC r})=u(y)\exp(i\KK\cdot{\VEC r}_\parallel)$, 
so the first (nonrelativistic) term vanishes at $\kk=0$, and the 
splitting is solely due to the relativistic 
term of Eq.~(\ref{relvel}).
This is not the case in real crystals: here it is the {\em nonrelativistic} 
term that gives the major contribution to the slope at $\bar\Gamma$. This 
is illustrated in Fig.~\ref{real} by the examples of three well-studied 
hexagonal surfaces: the classical case of Au(111), the surface alloy Bi/Ag(111) 
exhibiting giant splitting, and the topological insulator Bi$_2$Se$_3$. 
In the upper row of Fig.~\ref{real} the numerical derivative of the 
eigenenergy $dE(\kk)/d\kk$ (total velocity) is compared with the 
nonrelativistic (classical) velocity $v_{\text c}=2\bra{\psi}\hat p_\parallel\ket{\psi}$. 

The wave functions $\psi$ were calculated with the Hamiltonian of      
\eq{sreq} with linear augmented plane wave method in a repeated slab   
geometry. The density-functional-theory calculations                   
\cite{*[{The self-consistent potential is obtained in local density 
approximation with the full-potential linear augmented plane waves method,
see }] [{}] KSS99} 
were performed for symmetric slabs (19 layers for noble metals, 
\cite{*[{Properties of the surface states in noble metals 
are addressed in detail in }] [{}] Lobo-Checa2011,*Borghetti2012}
9 for Bi/Ag(111),
\footnote{For the crystal structure of the Ag$_2$Bi surface alloy 
and the properties of the surface state see, e.g., 
Refs.~\onlinecite{Meier2008,Gierz2010,Ast2007,Ast2008}} 
and 25 for Bi$_2$Se$_3$
\cite{*[{For the electronic structure of ${\mathrm{Bi}}_{2}{\mathrm{Se}}_{3}$ 
see, e.g., }] [{}] Kim2011}),
and a small perturbation at one of the surfaces was introduced afterwards 
to disentangle the surface states at the opposite surfaces. Relativistic 
effects are treated within the second-variational two-component Koelling-Harmon
approximation.~\cite{KOH77,MCD80} The potential gradient is taken into 
account only in the muffin-tin spheres, where the spin-orbit term takes 
the form $\beta[V'_r(r)/r]\,\boldsymbol\sigma\cdot\hat{\mathbf L}$. Its 
contribution to the group velocity is then given by the expectation value 
$v_{\text r}=\bra{\psi}\beta\,\boldsymbol\sigma\cdot[\VEC{r}\times 
\boldsymbol\tau_\parallel]\,V'_r(r)/r\ket{\psi}$.

The velocities calculated by this formula are shown in the middle row of 
Fig.~\ref{real}. The main message is that at $\bar\Gamma$ the spin-orbit 
part is an order of magnitude smaller than the full velocity (circles in the 
upper row). This result is corroborated by the calculation of the classical 
velocity $v_{\text c}$ (solid lines in the upper row), which is seen to be 
very close to the full velocity (found by numerical differentiation). 
Clearly, the splitting here follows a strikingly different scenario from the one 
of the standard Rashba model: crucial is the influence of the relativistic term 
on the wave function, which then gives rise to the strong splitting through the 
{\em purely nonrelativistic} velocity operator. In other words, a perturbation 
theory based on non- or scalar-relativistic wave functions, which in the Rashba 
model is exact at $\bar\Gamma$, is here manifestly inapplicable. The parameter  
$\alpha_{\text R}$ that can be derived by fitting experimental or {\it ab initio} 
energies to \eq{rbh}, is, thus, not the measure of either normal or in-plane 
gradient (or the respective charge  density asymmetry). Moreover, the true 
Rashba parameters in the three crystals (see the middle row of Fig.~\ref{real}) 
differ much less than the full slope at $\bar\Gamma$ (upper row). The dependence 
of the velocity ingredients on the atomic number can be inferred from 
Table~\ref{tab}, which lists also Cu and Ag. In the series Cu, Ag, Au, 
the relativistic term $v_{\text r}$ steadily grows, and it is reasonably 
large for Bi based crystals -- fully consistent with the Rashba model. 
However, there is no apparent correlation between $v_{\text r}$ and the 
ultimate effect.

\begin{table}[b]
\caption{\label{tab} Group velocity at $\bar\Gamma$ for the surface states 
at some (111) surfaces. Values obtained by numerical differentiation (upper 
row) are compared to nonrelativistic and spin-orbit parts obtained from the 
wave functions. Identity \eq{relvel} is satisfied only with a certain accuracy 
due to the variational character of the wave functions. 
\cite{*[{The uncertainty in $v_{\text c}$ as large as 0.1~Ry$\cdot$a.u. has
been reported, e.g., for bulk states in Be in }] [{. The method of second 
variation for the spin-orbit coupling \cite{KOH77} introduces additional
error because it neglects the dependence of the radial basis functions
on the relativistic quantum number $\kappa$ \cite{MCD80}}] Krasovskii1993}
}
\begin{ruledtabular}
\begin{tabular}{rlllll}
                 & Cu    & Ag    & Au    & Ag$_2$Bi  & Bi$_2$Se$_3$ \\ 
 $dE(\kk)/d\kk$  & 0.02  & 0.01  & 0.10  & 0.32      & 0.26         \\ 
 nonrelativistic  $v_{\text c}$ & 0.02  & 0.02  & 0.12  & 0.32      & 0.29         \\ 
    relativistic  $v_{\text r}$ & 0.002 & 0.006 & 0.014 & 0.022     & 0.010        \\  
\end{tabular}
\end{ruledtabular}
\end{table}

\begin{figure}[t]  
\includegraphics[height=0.17\textheight,clip=true,trim=125 50 95 20]{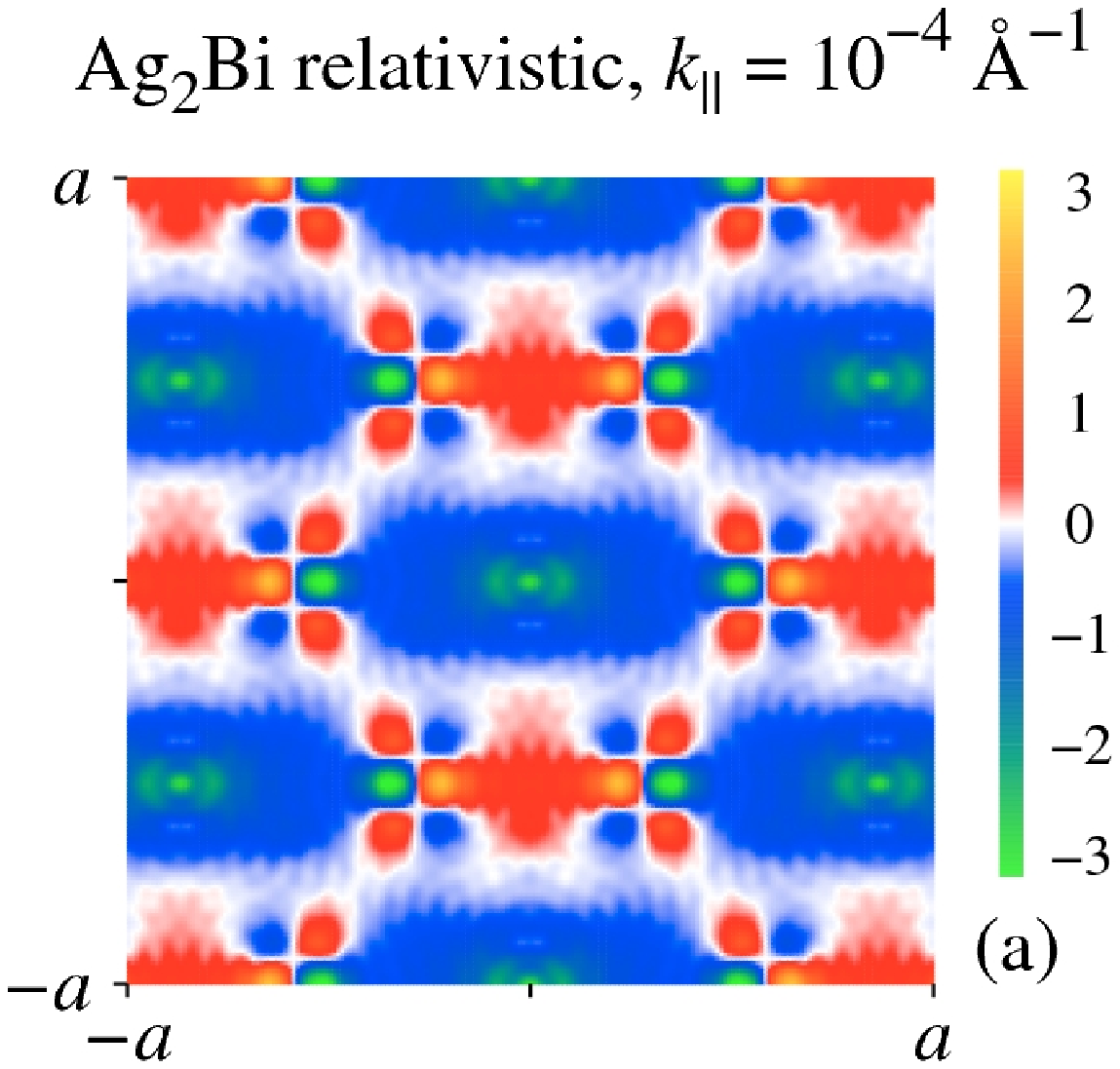}%
\includegraphics[height=0.17\textheight,clip=true,trim=125 50 80 20]{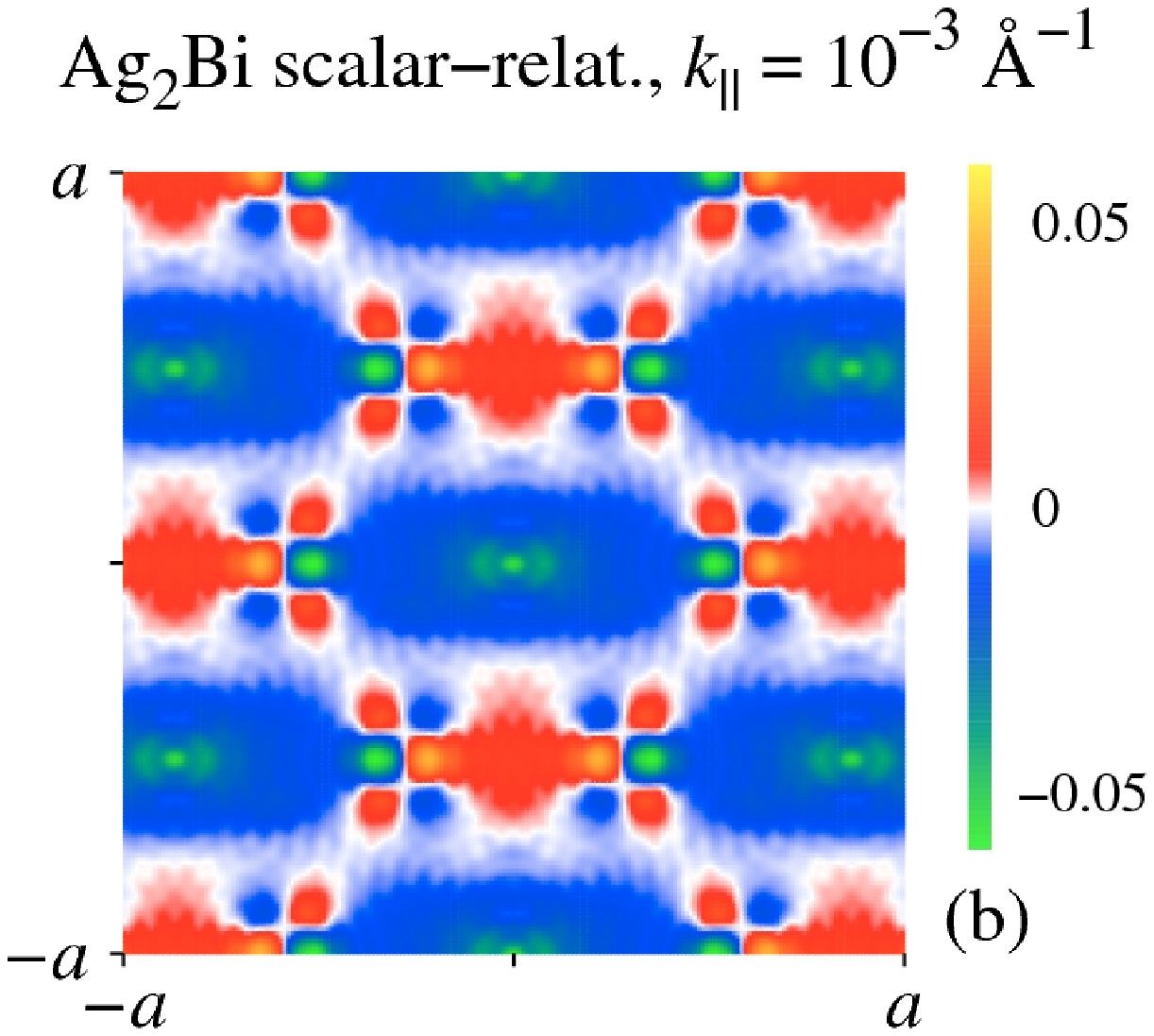}
\caption{\label{cur} (Color online) In-plane distribution of the $\KK$ 
projection of the classical current density $j(\VEC{r}_\parallel)$ (in the 
units of Ry$\cdot$a.u.) for the surface state at Bi/Ag(111) with $\KK$ along 
the horizontal axis. 
(a) relativistic $j(\VEC{r}_\parallel)$ map for $\kk=10^{-4}$\AA$^{-1}$
for the inner-circle surface state, i.e., for the one with the smaller
$\kk$ at a given $E$, hence the negative group velocity. 
(b) scalar relativistic map for $\kk=10^{-3}$\AA$^{-1}$.
}
\end{figure}

In order to understand how the spin-orbit term modifies the wave function, 
let us consider the classical current density
$\VEC{j}(\VEC{r})=2\text{Re}\,\psi(\VEC{r})^*[-i\boldsymbol\nabla]\psi(\VEC{r})$.
It is instructive to visualize the lateral spatial structure of the surface 
state by plotting the projection of $\VEC{j}$ in the direction of motion, 
$j=\VEC{j}\cdot\boldsymbol\tau_\parallel$. After averaging along the surface 
normal we obtain a scalar field $j(\VEC{r}_\parallel)$, which is shown by the 
color maps in Fig.~\ref{cur}. The average value of $j(\VEC{r}_\parallel)$ over 
the 2D unit cell is the classical velocity $v_{\text c}$. (Here $j$ refers only 
to the periodic part of the Bloch function: the trivial spatially constant 
term $2\kk$ is dropped.)
Let us compare the current density distribution in relativistic and scalar 
relativistic calculations at $\kk\to 0$. Figure~\ref{cur} shows an example 
for Bi/Ag(111): the relativistic $j(\VEC{r}_\parallel)$ map is for 
$\kk=10^{-4}$\AA$^{-1}$ and scalar relativistic for $10^{-3}$\AA$^{-1}$. The 
two maps are seen to be identical down to tiniest details, with the only 
difference being the scale: the relativistic amplitude at the smaller $\kk$ is 
everywhere 50 times the scalar relativistic one at the larger $\kk$. Generally, 
with, as well as without spin-orbit coupling, the function $j(\VEC{r}_\parallel)$ 
retains its shape up to rather large $\kk$, and in the scalar relativistic case 
its amplitude steadily grows with $\kk$. In the relativistic case the shape is 
the same, but the amplitude is large already at $\bar\Gamma$. This same behavior 
of the current density is demonstrated also by noble metals.

The above current-density considerations offer a rather transparent picture of how 
the spin-orbit interaction affects the surface state: its principal role is to modify 
the wave function in such a way that the function $j(\VEC{r}_\parallel)$ acquire a 
large amplitude at $\bar\Gamma$ (which in the scalar relativistic case would happen 
at large $\kk$). To put it most concisely, the spin-orbit term shifts the distribution
$j(\KK;\VEC{r}_\parallel)$ along $\KK$. This relativistic effect is omitted
in the Rashba model because there it is $j(\VEC{r}_\parallel)=0$.

\section{conclusions}
We can conclude that the relativistic splitting of surface states has the
same microscopic nature in all the materials, regardless of its size or
topology (trivial surface states in the metals or topologically protected 
Dirac cone in Bi$_2$Se$_3$). In particular, Au(111) turns out to be not a 
paradigm case of an RB system, which explains the recently encountered 
``departure from the Rashba model'' in the spin-flip excitations at 
Au(111).~\cite{Azpiroz2013} Thus, neither the parabolicity of $E(\kk)$ nor 
the full in-plane spin polarization of the states are an indication of the 
RB scenario. The relativistic wave function modification revealed by 
Figs.~\ref{real} and \ref{cur} has important implications in a wide range of 
scattering and excitation processes: this property should be taken into account 
in modeling the inelastic scattering and lifetime effects~\cite{Nechaev2010} 
and especially electron and spin transport by the surface states, which is 
directly connected to the current density distribution. Furthermore, the 
spatial structure of the wave function reflects itself in angle-resolved 
photoemission: the component of the electric field of the incident light 
along $\KK$ emphasizes the relevant feature of the initial state via the 
matrix element of the classical velocity operator 
$\bra{\Phi_{\text{final}}}\hat p_\parallel\ket{\psi}$. According to the 
present theory the $\kk$ dependence of the matrix element would be different 
for the splitting of the RB type and of the present type.
  
To summarize, in order to manipulate the spin-orbit splitting it is necessary 
to know how it arises. The present study has established that in crystals built 
of high-Z atoms the splitting comes primarily from a relativistic modification 
of the wave function that makes the {\em classical} current carried by the surface 
state finite at $\bar\Gamma$. This spin-orbit induced transformation can be 
described as a shift of the current density distribution along $\KK$. This is 
by contrast to the Rashba model, in which the spin-orbit interaction does not 
modify the in-plane structure of the wave function. The discovered scenario is 
followed equally closely by noble metals, surface alloys, and topological 
insulators irrespective of the strength of the effect and topological nature.

\begin{acknowledgments}
The author is grateful to R.~Kuzian, I.~Nechaev, and I.~Tokatly for the 
critical reading of this manuscript and many useful discussions. Partial 
support is acknowledged from the Spanish Ministerio de Ciencia e 
Innovaci\'on (Grant No. FIS2010-19609-C02-02).
  \end{acknowledgments}


\begin{thebibliography}{10}%
\makeatletter
\providecommand \@ifxundefined [1]{%
 \ifx #1\undefined \expandafter \@firstoftwo
 \else \expandafter \@secondoftwo
\fi
}%
\providecommand \@ifnum [1]{%
 \ifnum #1\expandafter \@firstoftwo
 \else \expandafter \@secondoftwo
\fi
}%
\providecommand \enquote [1]{``#1''}%
\providecommand \bibnamefont  [1]{#1}%
\providecommand \bibfnamefont [1]{#1}%
\providecommand \citenamefont [1]{#1}%
\providecommand\href[0]{\@sanitize\@href}%
\providecommand\@href[1]{\endgroup\@@startlink{#1}\endgroup\@@href}%
\providecommand\@@href[1]{#1\@@endlink}%
\providecommand \@sanitize [0]{\begingroup\catcode`\&12\catcode`\#12\relax}%
\@ifxundefined \pdfoutput {\@firstoftwo}{%
 \@ifnum{\z@=\pdfoutput}{\@firstoftwo}{\@secondoftwo}%
}{%
 \providecommand\@@startlink[1]{\leavevmode}%
 \providecommand\@@endlink[0]{}%
}{%
 \providecommand\@@startlink[1]{%
  \leavevmode
  \pdfstartlink
   attr{/Border[0 0 1 ]/H/I/C[0 1 1]}%
   user{/Subtype/Link/A<</Type/Action/S/URI/URI(#1)>>}%
  \relax
 }%
 \providecommand\@@endlink[0]{\pdfendlink}%
}%
\providecommand \url  [0]{\begingroup\@sanitize \@url }%
\providecommand \@url [1]{\endgroup\@href {#1}{\urlprefix}}%
\providecommand \urlprefix [0]{URL }%
\providecommand \Eprint[0]{\href }%
\@ifxundefined \urlstyle {%
  \providecommand \doi [1]{doi:\discretionary{}{}{}#1}%
}{%
  \providecommand \doi [0]{doi:\discretionary{}{}{}\begingroup
  \urlstyle{rm}\Url }%
}%
\providecommand \doibase [0]{http://dx.doi.org/}%
\providecommand \Doi[1]{\href{\doibase#1}}%
\providecommand \bibAnnote [3]{%
  \BibitemShut{#1}%
  \begin{quotation}\noindent
    \textsc{Key:}\ #2\\\textsc{Annotation:}\ #3%
  \end{quotation}%
}%
\providecommand \bibAnnoteFile [2]{%
  \IfFileExists{#2}{\bibAnnote {#1} {#2} {\input{#2}}}{}%
}%
\providecommand \typeout [0]{\immediate \write \m@ne }%
\providecommand \selectlanguage [0]{\@gobble}%
\providecommand \bibinfo [0]{\@secondoftwo}%
\providecommand \bibfield [0]{\@secondoftwo}%
\providecommand \translation [1]{[#1]}%
\providecommand \BibitemOpen[0]{}%
\providecommand \bibitemStop [0]{}%
\providecommand \bibitemNoStop [0]{.\EOS\space}%
\providecommand \EOS [0]{\spacefactor3000\relax}%
\providecommand \BibitemShut [1]{\csname bibitem#1\endcsname}%
\bibitem{Wolf_science2001}%
  \BibitemOpen
  \bibfield{author}{%
  \bibinfo {author} {\bibfnamefont{S.~A.}\ \bibnamefont{Wolf}}, \bibinfo
  {author} {\bibfnamefont{D.~D.}\ \bibnamefont{Awschalom}}, \bibinfo {author}
  {\bibfnamefont{R.~A.}\ \bibnamefont{Buhrman}}, \bibinfo {author}
  {\bibfnamefont{J.~M.}\ \bibnamefont{Daughton}}, \bibinfo {author}
  {\bibfnamefont{S.}~\bibnamefont{von Molnár}}, \bibinfo {author}
  {\bibfnamefont{M.~L.}\ \bibnamefont{Roukes}}, \bibinfo {author}
  {\bibfnamefont{A.~Y.}\ \bibnamefont{Chtchelkanova}},\ and\ \bibinfo {author}
  {\bibfnamefont{D.~M.}\ \bibnamefont{Treger}},\ }%
  \bibfield{journal}{%
  \Doi{10.1126/science.1065389}{\bibinfo {journal} {Science}}\ }%
  \textbf{\bibinfo {volume} {294}},\ \bibinfo {pages} {1488} (\bibinfo {year}
  {2001})%
  \bibAnnoteFile{NoStop}{Wolf_science2001}%
\bibitem{zutic2004}%
  \BibitemOpen
  \bibfield{author}{%
  \bibinfo {author} {\bibfnamefont{I.}~\bibnamefont{\ifmmode \check{Z}\else
  \v{Z}\fi{}uti\ifmmode~\acute{c}\else \'{c}\fi{}}}, \bibinfo {author}
  {\bibfnamefont{J.}~\bibnamefont{Fabian}},\ and\ \bibinfo {author}
  {\bibfnamefont{S.}~\bibnamefont{Das~Sarma}},\ }%
  \bibfield{journal}{%
  \Doi{10.1103/RevModPhys.76.323}{\bibinfo {journal} {Rev. Mod. Phys.}}\ }%
  \textbf{\bibinfo {volume} {76}},\ \bibinfo {pages} {323} (\bibinfo {year}
  {2004})%
  \bibAnnoteFile{NoStop}{zutic2004}%
\bibitem{RB1984JL}%
  \BibitemOpen
  \bibfield{author}{%
  \bibinfo {author} {\bibfnamefont{Y.~A.}\ \bibnamefont{Bychkov}}\ and\
  \bibinfo {author} {\bibfnamefont{E.~I.}\ \bibnamefont{Rashba}},\ }%
  \bibfield{journal}{%
  \bibinfo {journal} {JETP Lett.}\ }%
  \textbf{\bibinfo {volume} {39}},\ \bibinfo {pages} {78} (\bibinfo {year}
  {1984})%
  \bibAnnoteFile{NoStop}{RB1984JL}%
\bibitem{RB1984JPh}%
  \BibitemOpen
  \bibfield{author}{%
  \bibinfo {author} {\bibfnamefont{Y.~A.}\ \bibnamefont{Bychkov}}\ and\
  \bibinfo {author} {\bibfnamefont{E.~I.}\ \bibnamefont{Rashba}},\ }%
  \bibfield{journal}{%
  \bibinfo {journal} {J. Phys. C}\ }%
  \textbf{\bibinfo {volume} {17}},\ \bibinfo {pages} {6039} (\bibinfo {year}
  {1984})%
  \bibAnnoteFile{NoStop}{RB1984JPh}%
\bibitem{DD1990}%
  \BibitemOpen
  \bibfield{author}{%
  \bibinfo {author} {\bibfnamefont{S.}~\bibnamefont{Datta}}\ and\ \bibinfo
  {author} {\bibfnamefont{B.}~\bibnamefont{Das}},\ }%
  \bibfield{journal}{%
  \bibinfo {journal} {Appl. Phys. Lett.}\ }%
  \textbf{\bibinfo {volume} {56}},\ \bibinfo {pages} {665} (\bibinfo {year}
  {1990})%
  \bibAnnoteFile{NoStop}{DD1990}%
\bibitem{Koga2002}%
  \BibitemOpen
  \bibfield{author}{%
  \bibinfo {author} {\bibfnamefont{T.}~\bibnamefont{Koga}}, \bibinfo {author}
  {\bibfnamefont{J.}~\bibnamefont{Nitta}}, \bibinfo {author}
  {\bibfnamefont{H.}~\bibnamefont{Takayanagi}},\ and\ \bibinfo {author}
  {\bibfnamefont{S.}~\bibnamefont{Datta}},\ }%
  \bibfield{journal}{%
  \Doi{10.1103/PhysRevLett.88.126601}{\bibinfo {journal} {Phys. Rev. Lett.}}\
  }%
  \textbf{\bibinfo {volume} {88}},\ \bibinfo {pages} {126601} (\bibinfo {year}
  {2002})%
  \bibAnnoteFile{NoStop}{Koga2002}%
\bibitem{LaShell1996}%
  \BibitemOpen
  \bibfield{author}{%
  \bibinfo {author} {\bibfnamefont{S.}~\bibnamefont{LaShell}}, \bibinfo
  {author} {\bibfnamefont{B.~A.}\ \bibnamefont{McDougall}},\ and\ \bibinfo
  {author} {\bibfnamefont{E.}~\bibnamefont{Jensen}},\ }%
  \bibfield{journal}{%
  \Doi{10.1103/PhysRevLett.77.3419}{\bibinfo {journal} {Phys. Rev. Lett.}}\ }%
  \textbf{\bibinfo {volume} {77}},\ \bibinfo {pages} {3419} (\bibinfo {year}
  {1996})%
  \bibAnnoteFile{NoStop}{LaShell1996}%
\bibitem{Nicolay2001}%
  \BibitemOpen
  \bibfield{author}{%
  \bibinfo {author} {\bibfnamefont{G.}~\bibnamefont{Nicolay}}, \bibinfo
  {author} {\bibfnamefont{F.}~\bibnamefont{Reinert}}, \bibinfo {author}
  {\bibfnamefont{S.}~\bibnamefont{H\"ufner}},\ and\ \bibinfo {author}
  {\bibfnamefont{P.}~\bibnamefont{Blaha}},\ }%
  \bibfield{journal}{%
  \Doi{10.1103/PhysRevB.65.033407}{\bibinfo {journal} {Phys. Rev. B}}\ }%
  \textbf{\bibinfo {volume} {65}},\ \bibinfo {pages} {033407} (\bibinfo {year}
  {2001})%
  \bibAnnoteFile{NoStop}{Nicolay2001}%
\bibitem{Wissing2013}%
  \BibitemOpen
  \bibfield{author}{%
  \bibinfo {author} {\bibfnamefont{S.~N.~P.}\ \bibnamefont{Wissing}}, \bibinfo
  {author} {\bibfnamefont{C.}~\bibnamefont{Eibl}}, \bibinfo {author}
  {\bibfnamefont{A.}~\bibnamefont{Zumb\"ulte}}, \bibinfo {author}
  {\bibfnamefont{A.~B.}\ \bibnamefont{Schmidt}}, \bibinfo {author}
  {\bibfnamefont{J.}~\bibnamefont{Braun}}, \bibinfo {author}
  {\bibfnamefont{J.}~\bibnamefont{Min\'ar}}, \bibinfo {author}
  {\bibfnamefont{H.}~\bibnamefont{Ebert}},\ and\ \bibinfo {author}
  {\bibfnamefont{M.}~\bibnamefont{Donath}},\ }%
  \bibfield{journal}{%
  \bibinfo {journal} {New J. Phys.}\ }%
  \textbf{\bibinfo {volume} {15}},\ \bibinfo {pages} {105001} (\bibinfo {year}
  {2013})%
  \bibAnnoteFile{NoStop}{Wissing2013}%
\bibitem{Rotenberg1998}%
  \BibitemOpen
  \bibfield{author}{%
  \bibinfo {author} {\bibfnamefont{E.}~\bibnamefont{Rotenberg}}\ and\ \bibinfo
  {author} {\bibfnamefont{S.~D.}\ \bibnamefont{Kevan}},\ }%
  \bibfield{journal}{%
  \Doi{10.1103/PhysRevLett.80.2905}{\bibinfo {journal} {Phys. Rev. Lett.}}\ }%
  \textbf{\bibinfo {volume} {80}},\ \bibinfo {pages} {2905} (\bibinfo {year}
  {1998})%
  \bibAnnoteFile{NoStop}{Rotenberg1998}%
\bibitem{Rotenberg1999}%
  \BibitemOpen
  \bibfield{author}{%
  \bibinfo {author} {\bibfnamefont{E.}~\bibnamefont{Rotenberg}}, \bibinfo
  {author} {\bibfnamefont{J.~W.}\ \bibnamefont{Chung}},\ and\ \bibinfo {author}
  {\bibfnamefont{S.~D.}\ \bibnamefont{Kevan}},\ }%
  \bibfield{journal}{%
  \Doi{10.1103/PhysRevLett.82.4066}{\bibinfo {journal} {Phys. Rev. Lett.}}\ }%
  \textbf{\bibinfo {volume} {82}},\ \bibinfo {pages} {4066} (\bibinfo {year}
  {1999})%
  \bibAnnoteFile{NoStop}{Rotenberg1999}%
\bibitem{Hochstrasser2002}%
  \BibitemOpen
  \bibfield{author}{%
  \bibinfo {author} {\bibfnamefont{M.}~\bibnamefont{Hochstrasser}}, \bibinfo
  {author} {\bibfnamefont{J.~G.}\ \bibnamefont{Tobin}}, \bibinfo {author}
  {\bibfnamefont{E.}~\bibnamefont{Rotenberg}},\ and\ \bibinfo {author}
  {\bibfnamefont{S.~D.}\ \bibnamefont{Kevan}},\ }%
  \bibfield{journal}{%
  \Doi{10.1103/PhysRevLett.89.216802}{\bibinfo {journal} {Phys. Rev. Lett.}}\
  }%
  \textbf{\bibinfo {volume} {89}},\ \bibinfo {pages} {216802} (\bibinfo {year}
  {2002})%
  \bibAnnoteFile{NoStop}{Hochstrasser2002}%
\bibitem{Koitzsch2005}%
  \BibitemOpen
  \bibfield{author}{%
  \bibinfo {author} {\bibfnamefont{C.}~\bibnamefont{Koitzsch}}, \bibinfo
  {author} {\bibfnamefont{C.}~\bibnamefont{Battaglia}}, \bibinfo {author}
  {\bibfnamefont{F.}~\bibnamefont{Clerc}}, \bibinfo {author}
  {\bibfnamefont{L.}~\bibnamefont{Despont}}, \bibinfo {author}
  {\bibfnamefont{M.~G.}\ \bibnamefont{Garnier}},\ and\ \bibinfo {author}
  {\bibfnamefont{P.}~\bibnamefont{Aebi}},\ }%
  \bibfield{journal}{%
  \Doi{10.1103/PhysRevLett.95.126401}{\bibinfo {journal} {Phys. Rev. Lett.}}\
  }%
  \textbf{\bibinfo {volume} {95}},\ \bibinfo {pages} {126401} (\bibinfo {year}
  {2005})%
  \bibAnnoteFile{NoStop}{Koitzsch2005}%
\bibitem{Schiller2005}%
  \BibitemOpen
  \bibfield{author}{%
  \bibinfo {author} {\bibfnamefont{F.}~\bibnamefont{Schiller}}, \bibinfo
  {author} {\bibfnamefont{R.}~\bibnamefont{Keyling}}, \bibinfo {author}
  {\bibfnamefont{E.~V.}\ \bibnamefont{Chulkov}},\ and\ \bibinfo {author}
  {\bibfnamefont{J.~E.}\ \bibnamefont{Ortega}},\ }%
  \bibfield{journal}{%
  \Doi{10.1103/PhysRevLett.95.126402}{\bibinfo {journal} {Phys. Rev. Lett.}}\
  }%
  \textbf{\bibinfo {volume} {95}},\ \bibinfo {pages} {126402} (\bibinfo {year}
  {2005})%
  \bibAnnoteFile{NoStop}{Schiller2005}%
\bibitem{Shikin2008}%
  \BibitemOpen
  \bibfield{author}{%
  \bibinfo {author} {\bibfnamefont{A.~M.}\ \bibnamefont{Shikin}}, \bibinfo
  {author} {\bibfnamefont{A.}~\bibnamefont{Varykhalov}}, \bibinfo {author}
  {\bibfnamefont{G.~V.}\ \bibnamefont{Prudnikova}}, \bibinfo {author}
  {\bibfnamefont{D.}~\bibnamefont{Usachov}}, \bibinfo {author}
  {\bibfnamefont{V.~K.}\ \bibnamefont{Adamchuk}}, \bibinfo {author}
  {\bibfnamefont{Y.}~\bibnamefont{Yamada}}, \bibinfo {author}
  {\bibfnamefont{J.~D.}\ \bibnamefont{Riley}},\ and\ \bibinfo {author}
  {\bibfnamefont{O.}~\bibnamefont{Rader}},\ }%
  \bibfield{journal}{%
  \Doi{10.1103/PhysRevLett.100.057601}{\bibinfo {journal} {Phys. Rev. Lett.}}\
  }%
  \textbf{\bibinfo {volume} {100}},\ \bibinfo {pages} {057601} (\bibinfo {year}
  {2008})%
  \bibAnnoteFile{NoStop}{Shikin2008}%
\bibitem{Rybkin2012}%
  \BibitemOpen
  \bibfield{author}{%
  \bibinfo {author} {\bibfnamefont{A.~G.}\ \bibnamefont{Rybkin}}, \bibinfo
  {author} {\bibfnamefont{E.~E.}\ \bibnamefont{Krasovskii}}, \bibinfo {author}
  {\bibfnamefont{D.}~\bibnamefont{Marchenko}}, \bibinfo {author}
  {\bibfnamefont{E.~V.}\ \bibnamefont{Chulkov}}, \bibinfo {author}
  {\bibfnamefont{A.}~\bibnamefont{Varykhalov}}, \bibinfo {author}
  {\bibfnamefont{O.}~\bibnamefont{Rader}},\ and\ \bibinfo {author}
  {\bibfnamefont{A.~M.}\ \bibnamefont{Shikin}},\ }%
  \bibfield{journal}{%
  \Doi{10.1103/PhysRevB.86.035117}{\bibinfo {journal} {Phys. Rev. B}}\ }%
  \textbf{\bibinfo {volume} {86}},\ \bibinfo {pages} {035117} (\bibinfo {year}
  {2012})%
  \bibAnnoteFile{NoStop}{Rybkin2012}%
\bibitem{Agergaard2001}%
  \BibitemOpen
  \bibfield{author}{%
  \bibinfo {author} {\bibfnamefont{S.}~\bibnamefont{Agergaard}}, \bibinfo
  {author} {\bibfnamefont{C.}~\bibnamefont{S{\o}ndergaard}}, \bibinfo {author}
  {\bibfnamefont{H.}~\bibnamefont{Li}}, \bibinfo {author}
  {\bibfnamefont{M.~B.}\ \bibnamefont{Nielsen}}, \bibinfo {author}
  {\bibfnamefont{S.~V.}\ \bibnamefont{Hoffmann}}, \bibinfo {author}
  {\bibfnamefont{Z.}~\bibnamefont{Li}},\ and\ \bibinfo {author}
  {\bibfnamefont{P.}~\bibnamefont{Hofmann}},\ }%
  \bibfield{journal}{%
  \bibinfo {journal} {New J. Phys.}\ }%
  \textbf{\bibinfo {volume} {3}},\ \bibinfo {pages} {15} (\bibinfo {year}
  {2001})%
  \bibAnnoteFile{NoStop}{Agergaard2001}%
\bibitem{Koroteev2004}%
  \BibitemOpen
  \bibfield{author}{%
  \bibinfo {author} {\bibfnamefont{Y.~M.}\ \bibnamefont{Koroteev}}, \bibinfo
  {author} {\bibfnamefont{G.}~\bibnamefont{Bihlmayer}}, \bibinfo {author}
  {\bibfnamefont{J.~E.}\ \bibnamefont{Gayone}}, \bibinfo {author}
  {\bibfnamefont{E.~V.}\ \bibnamefont{Chulkov}}, \bibinfo {author}
  {\bibfnamefont{S.}~\bibnamefont{Bl\"ugel}}, \bibinfo {author}
  {\bibfnamefont{P.~M.}\ \bibnamefont{Echenique}},\ and\ \bibinfo {author}
  {\bibfnamefont{P.}~\bibnamefont{Hofmann}},\ }%
  \bibfield{journal}{%
  \Doi{10.1103/PhysRevLett.93.046403}{\bibinfo {journal} {Phys. Rev. Lett.}}\
  }%
  \textbf{\bibinfo {volume} {93}},\ \bibinfo {pages} {046403} (\bibinfo {year}
  {2004})%
  \bibAnnoteFile{NoStop}{Koroteev2004}%
\bibitem{Sugawara2006}%
  \BibitemOpen
  \bibfield{author}{%
  \bibinfo {author} {\bibfnamefont{K.}~\bibnamefont{Sugawara}}, \bibinfo
  {author} {\bibfnamefont{T.}~\bibnamefont{Sato}}, \bibinfo {author}
  {\bibfnamefont{S.}~\bibnamefont{Souma}}, \bibinfo {author}
  {\bibfnamefont{T.}~\bibnamefont{Takahashi}}, \bibinfo {author}
  {\bibfnamefont{M.}~\bibnamefont{Arai}},\ and\ \bibinfo {author}
  {\bibfnamefont{T.}~\bibnamefont{Sasaki}},\ }%
  \bibfield{journal}{%
  \Doi{10.1103/PhysRevLett.96.046411}{\bibinfo {journal} {Phys. Rev. Lett.}}\
  }%
  \textbf{\bibinfo {volume} {96}},\ \bibinfo {pages} {046411} (\bibinfo {year}
  {2006})%
  \bibAnnoteFile{NoStop}{Sugawara2006}%
\bibitem{Kimura2010}%
  \BibitemOpen
  \bibfield{author}{%
  \bibinfo {author} {\bibfnamefont{A.}~\bibnamefont{Kimura}}, \bibinfo {author}
  {\bibfnamefont{E.~E.}\ \bibnamefont{Krasovskii}}, \bibinfo {author}
  {\bibfnamefont{R.}~\bibnamefont{Nishimura}}, \bibinfo {author}
  {\bibfnamefont{K.}~\bibnamefont{Miyamoto}}, \bibinfo {author}
  {\bibfnamefont{T.}~\bibnamefont{Kadono}}, \bibinfo {author}
  {\bibfnamefont{K.}~\bibnamefont{Kanomaru}}, \bibinfo {author}
  {\bibfnamefont{E.~V.}\ \bibnamefont{Chulkov}}, \bibinfo {author}
  {\bibfnamefont{G.}~\bibnamefont{Bihlmayer}}, \bibinfo {author}
  {\bibfnamefont{K.}~\bibnamefont{Shimada}}, \bibinfo {author}
  {\bibfnamefont{H.}~\bibnamefont{Namatame}},\ and\ \bibinfo {author}
  {\bibfnamefont{M.}~\bibnamefont{Taniguchi}},\ }%
  \bibfield{journal}{%
  \Doi{10.1103/PhysRevLett.105.076804}{\bibinfo {journal} {Phys. Rev. Lett.}}\
  }%
  \textbf{\bibinfo {volume} {105}},\ \bibinfo {pages} {076804} (\bibinfo {year}
  {2010})%
  \bibAnnoteFile{NoStop}{Kimura2010}%
\bibitem{Meier2008}%
  \BibitemOpen
  \bibfield{author}{%
  \bibinfo {author} {\bibfnamefont{F.}~\bibnamefont{Meier}}, \bibinfo {author}
  {\bibfnamefont{H.}~\bibnamefont{Dil}}, \bibinfo {author}
  {\bibfnamefont{J.}~\bibnamefont{Lobo-Checa}}, \bibinfo {author}
  {\bibfnamefont{L.}~\bibnamefont{Patthey}},\ and\ \bibinfo {author}
  {\bibfnamefont{J.}~\bibnamefont{Osterwalder}},\ }%
  \bibfield{journal}{%
  \Doi{10.1103/PhysRevB.77.165431}{\bibinfo {journal} {Phys. Rev. B}}\ }%
  \textbf{\bibinfo {volume} {77}},\ \bibinfo {pages} {165431} (\bibinfo {year}
  {2008})%
  \bibAnnoteFile{NoStop}{Meier2008}%
\bibitem{Gierz2010}%
  \BibitemOpen
  \bibfield{author}{%
  \bibinfo {author} {\bibfnamefont{I.}~\bibnamefont{Gierz}}, \bibinfo {author}
  {\bibfnamefont{B.}~\bibnamefont{Stadtm\"uller}}, \bibinfo {author}
  {\bibfnamefont{J.}~\bibnamefont{Vuorinen}}, \bibinfo {author}
  {\bibfnamefont{M.}~\bibnamefont{Lindroos}}, \bibinfo {author}
  {\bibfnamefont{F.}~\bibnamefont{Meier}}, \bibinfo {author}
  {\bibfnamefont{J.~H.}\ \bibnamefont{Dil}}, \bibinfo {author}
  {\bibfnamefont{K.}~\bibnamefont{Kern}},\ and\ \bibinfo {author}
  {\bibfnamefont{C.~R.}\ \bibnamefont{Ast}},\ }%
  \bibfield{journal}{%
  \Doi{10.1103/PhysRevB.81.245430}{\bibinfo {journal} {Phys. Rev. B}}\ }%
  \textbf{\bibinfo {volume} {81}},\ \bibinfo {pages} {245430} (\bibinfo {year}
  {2010})%
  \bibAnnoteFile{NoStop}{Gierz2010}%
\bibitem{Ast2007}%
  \BibitemOpen
  \bibfield{author}{%
  \bibinfo {author} {\bibfnamefont{C.~R.}\ \bibnamefont{Ast}}, \bibinfo
  {author} {\bibfnamefont{J.}~\bibnamefont{Henk}}, \bibinfo {author}
  {\bibfnamefont{A.}~\bibnamefont{Ernst}}, \bibinfo {author}
  {\bibfnamefont{L.}~\bibnamefont{Moreschini}}, \bibinfo {author}
  {\bibfnamefont{M.~C.}\ \bibnamefont{Falub}}, \bibinfo {author}
  {\bibfnamefont{D.}~\bibnamefont{Pacil\'e}}, \bibinfo {author}
  {\bibfnamefont{P.}~\bibnamefont{Bruno}}, \bibinfo {author}
  {\bibfnamefont{K.}~\bibnamefont{Kern}},\ and\ \bibinfo {author}
  {\bibfnamefont{M.}~\bibnamefont{Grioni}},\ }%
  \bibfield{journal}{%
  \Doi{10.1103/PhysRevLett.98.186807}{\bibinfo {journal} {Phys. Rev. Lett.}}\
  }%
  \textbf{\bibinfo {volume} {98}},\ \bibinfo {pages} {186807} (\bibinfo {year}
  {2007})%
  \bibAnnoteFile{NoStop}{Ast2007}%
\bibitem{Ast2008}%
  \BibitemOpen
  \bibfield{author}{%
  \bibinfo {author} {\bibfnamefont{C.~R.}\ \bibnamefont{Ast}}, \bibinfo
  {author} {\bibfnamefont{D.}~\bibnamefont{Pacil\'e}}, \bibinfo {author}
  {\bibfnamefont{L.}~\bibnamefont{Moreschini}}, \bibinfo {author}
  {\bibfnamefont{M.~C.}\ \bibnamefont{Falub}}, \bibinfo {author}
  {\bibfnamefont{M.}~\bibnamefont{Papagno}}, \bibinfo {author}
  {\bibfnamefont{K.}~\bibnamefont{Kern}}, \bibinfo {author}
  {\bibfnamefont{M.}~\bibnamefont{Grioni}}, \bibinfo {author}
  {\bibfnamefont{J.}~\bibnamefont{Henk}}, \bibinfo {author}
  {\bibfnamefont{A.}~\bibnamefont{Ernst}}, \bibinfo {author}
  {\bibfnamefont{S.}~\bibnamefont{Ostanin}},\ and\ \bibinfo {author}
  {\bibfnamefont{P.}~\bibnamefont{Bruno}},\ }%
  \bibfield{journal}{%
  \Doi{10.1103/PhysRevB.77.081407}{\bibinfo {journal} {Phys. Rev. B}}\ }%
  \textbf{\bibinfo {volume} {77}},\ \bibinfo {pages} {081407} (\bibinfo {year}
  {2008})%
  \bibAnnoteFile{NoStop}{Ast2008}%
\bibitem{Gierz2009}%
  \BibitemOpen
  \bibfield{author}{%
  \bibinfo {author} {\bibfnamefont{I.}~\bibnamefont{Gierz}}, \bibinfo {author}
  {\bibfnamefont{T.}~\bibnamefont{Suzuki}}, \bibinfo {author}
  {\bibfnamefont{E.}~\bibnamefont{Frantzeskakis}}, \bibinfo {author}
  {\bibfnamefont{S.}~\bibnamefont{Pons}}, \bibinfo {author}
  {\bibfnamefont{S.}~\bibnamefont{Ostanin}}, \bibinfo {author}
  {\bibfnamefont{A.}~\bibnamefont{Ernst}}, \bibinfo {author}
  {\bibfnamefont{J.}~\bibnamefont{Henk}}, \bibinfo {author}
  {\bibfnamefont{M.}~\bibnamefont{Grioni}}, \bibinfo {author}
  {\bibfnamefont{K.}~\bibnamefont{Kern}},\ and\ \bibinfo {author}
  {\bibfnamefont{C.~R.}\ \bibnamefont{Ast}},\ }%
  \bibfield{journal}{%
  \Doi{10.1103/PhysRevLett.103.046803}{\bibinfo {journal} {Phys. Rev. Lett.}}\
  }%
  \textbf{\bibinfo {volume} {103}},\ \bibinfo {pages} {046803} (\bibinfo {year}
  {2009})%
  \bibAnnoteFile{NoStop}{Gierz2009}%
\bibitem{Sakamoto2009}%
  \BibitemOpen
  \bibfield{author}{%
  \bibinfo {author} {\bibfnamefont{K.}~\bibnamefont{Sakamoto}}, \bibinfo
  {author} {\bibfnamefont{T.}~\bibnamefont{Oda}}, \bibinfo {author}
  {\bibfnamefont{A.}~\bibnamefont{Kimura}}, \bibinfo {author}
  {\bibfnamefont{K.}~\bibnamefont{Miyamoto}}, \bibinfo {author}
  {\bibfnamefont{M.}~\bibnamefont{Tsujikawa}}, \bibinfo {author}
  {\bibfnamefont{A.}~\bibnamefont{Imai}}, \bibinfo {author}
  {\bibfnamefont{N.}~\bibnamefont{Ueno}}, \bibinfo {author}
  {\bibfnamefont{H.}~\bibnamefont{Namatame}}, \bibinfo {author}
  {\bibfnamefont{M.}~\bibnamefont{Taniguchi}}, \bibinfo {author}
  {\bibfnamefont{P.~E.~J.}\ \bibnamefont{Eriksson}},\ and\ \bibinfo {author}
  {\bibfnamefont{R.~I.~G.}\ \bibnamefont{Uhrberg}},\ }%
  \bibfield{journal}{%
  \Doi{10.1103/PhysRevLett.102.096805}{\bibinfo {journal} {Phys. Rev. Lett.}}\
  }%
  \textbf{\bibinfo {volume} {102}},\ \bibinfo {pages} {096805} (\bibinfo {year}
  {2009})%
  \bibAnnoteFile{NoStop}{Sakamoto2009}%
\bibitem{Hatta2009}%
  \BibitemOpen
  \bibfield{author}{%
  \bibinfo {author} {\bibfnamefont{S.}~\bibnamefont{Hatta}}, \bibinfo {author}
  {\bibfnamefont{T.}~\bibnamefont{Aruga}}, \bibinfo {author}
  {\bibfnamefont{Y.}~\bibnamefont{Ohtsubo}},\ and\ \bibinfo {author}
  {\bibfnamefont{H.}~\bibnamefont{Okuyama}},\ }%
  \bibfield{journal}{%
  \Doi{10.1103/PhysRevB.80.113309}{\bibinfo {journal} {Phys. Rev. B}}\ }%
  \textbf{\bibinfo {volume} {80}},\ \bibinfo {pages} {113309} (\bibinfo {year}
  {2009})%
  \bibAnnoteFile{NoStop}{Hatta2009}%
\bibitem{Yaji2010}%
  \BibitemOpen
  \bibfield{author}{%
  \bibinfo {author} {\bibfnamefont{K.}~\bibnamefont{Yaji}}, \bibinfo {author}
  {\bibfnamefont{Y.}~\bibnamefont{Ohtsubo}}, \bibinfo {author}
  {\bibfnamefont{S.}~\bibnamefont{Hatta}}, \bibinfo {author}
  {\bibfnamefont{H.}~\bibnamefont{Okuyama}}, \bibinfo {author}
  {\bibfnamefont{K.}~\bibnamefont{Miyamoto}}, \bibinfo {author}
  {\bibfnamefont{T.}~\bibnamefont{Okuda}}, \bibinfo {author}
  {\bibfnamefont{A.}~\bibnamefont{Kimura}}, \bibinfo {author}
  {\bibfnamefont{H.}~\bibnamefont{Namatame}}, \bibinfo {author}
  {\bibfnamefont{M.}~\bibnamefont{Taniguchi}},\ and\ \bibinfo {author}
  {\bibfnamefont{T.}~\bibnamefont{Aruga}},\ }%
  \bibfield{journal}{%
  \bibinfo {journal} {Nat. Commun.}\ }%
  \textbf{\bibinfo {volume} {1}},\ \bibinfo {pages} {17} (\bibinfo {year}
  {2010})%
  \bibAnnoteFile{NoStop}{Yaji2010}%
\bibitem{Ishizaka2011}%
  \BibitemOpen
  \bibfield{author}{%
  \bibinfo {author} {\bibfnamefont{K.}~\bibnamefont{Ishizaka}}, \bibinfo
  {author} {\bibfnamefont{M.~S.}\ \bibnamefont{Bahramy}}, \bibinfo {author}
  {\bibfnamefont{H.}~\bibnamefont{Murakawa}}, \bibinfo {author}
  {\bibfnamefont{M.}~\bibnamefont{Sakano}}, \bibinfo {author}
  {\bibfnamefont{T.}~\bibnamefont{Shimojima}}, \bibinfo {author}
  {\bibfnamefont{T.}~\bibnamefont{Sonobe}}, \bibinfo {author}
  {\bibfnamefont{K.}~\bibnamefont{Koizumi}}, \bibinfo {author}
  {\bibfnamefont{S.}~\bibnamefont{Shin}}, \bibinfo {author}
  {\bibfnamefont{H.}~\bibnamefont{Miyahara}}, \bibinfo {author}
  {\bibfnamefont{A.}~\bibnamefont{Kimura}}, \bibinfo {author}
  {\bibfnamefont{K.}~\bibnamefont{Miyamoto}}, \bibinfo {author}
  {\bibfnamefont{T.}~\bibnamefont{Okuda}}, \bibinfo {author}
  {\bibfnamefont{H.}~\bibnamefont{Namatame}}, \bibinfo {author}
  {\bibfnamefont{M.}~\bibnamefont{Taniguchi}}, \bibinfo {author}
  {\bibfnamefont{R.}~\bibnamefont{Arita}}, \bibinfo {author}
  {\bibfnamefont{N.}~\bibnamefont{Nagaosa}}, \bibinfo {author}
  {\bibfnamefont{K.}~\bibnamefont{Kobayashi}}, \bibinfo {author}
  {\bibfnamefont{Y.}~\bibnamefont{Murakami}}, \bibinfo {author}
  {\bibfnamefont{R.}~\bibnamefont{Kumai}}, \bibinfo {author}
  {\bibfnamefont{Y.}~\bibnamefont{Kaneko}}, \bibinfo {author}
  {\bibfnamefont{Y.}~\bibnamefont{Onose}},\ and\ \bibinfo {author}
  {\bibfnamefont{Y.}~\bibnamefont{Tokura}},\ }%
  \bibfield{journal}{%
  \Doi{10.1038/nmat3051}{\bibinfo {journal} {Nat. Mater.}}\ }%
  \textbf{\bibinfo {volume} {10}},\ \bibinfo {pages} {521} (\bibinfo {year}
  {2011})%
  \bibAnnoteFile{NoStop}{Ishizaka2011}%
\bibitem{Bahramy2011}%
  \BibitemOpen
  \bibfield{author}{%
  \bibinfo {author} {\bibfnamefont{M.~S.}\ \bibnamefont{Bahramy}}, \bibinfo
  {author} {\bibfnamefont{R.}~\bibnamefont{Arita}},\ and\ \bibinfo {author}
  {\bibfnamefont{N.}~\bibnamefont{Nagaosa}},\ }%
  \bibfield{journal}{%
  \Doi{10.1103/PhysRevB.84.041202}{\bibinfo {journal} {Phys. Rev. B}}\ }%
  \textbf{\bibinfo {volume} {84}},\ \bibinfo {pages} {041202} (\bibinfo {year}
  {2011})%
  \bibAnnoteFile{NoStop}{Bahramy2011}%
\bibitem{Landolt2012}%
  \BibitemOpen
  \bibfield{author}{%
  \bibinfo {author} {\bibfnamefont{G.}~\bibnamefont{Landolt}}, \bibinfo
  {author} {\bibfnamefont{S.~V.}\ \bibnamefont{Eremeev}}, \bibinfo {author}
  {\bibfnamefont{Y.~M.}\ \bibnamefont{Koroteev}}, \bibinfo {author}
  {\bibfnamefont{B.}~\bibnamefont{Slomski}}, \bibinfo {author}
  {\bibfnamefont{S.}~\bibnamefont{Muff}}, \bibinfo {author}
  {\bibfnamefont{T.}~\bibnamefont{Neupert}}, \bibinfo {author}
  {\bibfnamefont{M.}~\bibnamefont{Kobayashi}}, \bibinfo {author}
  {\bibfnamefont{V.~N.}\ \bibnamefont{Strocov}}, \bibinfo {author}
  {\bibfnamefont{T.}~\bibnamefont{Schmitt}}, \bibinfo {author}
  {\bibfnamefont{Z.~S.}\ \bibnamefont{Aliev}}, \bibinfo {author}
  {\bibfnamefont{M.~B.}\ \bibnamefont{Babanly}}, \bibinfo {author}
  {\bibfnamefont{I.~R.}\ \bibnamefont{Amiraslanov}}, \bibinfo {author}
  {\bibfnamefont{E.~V.}\ \bibnamefont{Chulkov}}, \bibinfo {author}
  {\bibfnamefont{J.}~\bibnamefont{Osterwalder}},\ and\ \bibinfo {author}
  {\bibfnamefont{J.~H.}\ \bibnamefont{Dil}},\ }%
  \bibfield{journal}{%
  \Doi{10.1103/PhysRevLett.109.116403}{\bibinfo {journal} {Phys. Rev. Lett.}}\
  }%
  \textbf{\bibinfo {volume} {109}},\ \bibinfo {pages} {116403} (\bibinfo {year}
  {2012})%
  \bibAnnoteFile{NoStop}{Landolt2012}%
\bibitem{Crepaldi2012}%
  \BibitemOpen
  \bibfield{author}{%
  \bibinfo {author} {\bibfnamefont{A.}~\bibnamefont{Crepaldi}}, \bibinfo
  {author} {\bibfnamefont{L.}~\bibnamefont{Moreschini}}, \bibinfo {author}
  {\bibfnamefont{G.}~\bibnamefont{Aut\`es}}, \bibinfo {author}
  {\bibfnamefont{C.}~\bibnamefont{Tournier-Colletta}}, \bibinfo {author}
  {\bibfnamefont{S.}~\bibnamefont{Moser}}, \bibinfo {author}
  {\bibfnamefont{N.}~\bibnamefont{Virk}}, \bibinfo {author}
  {\bibfnamefont{H.}~\bibnamefont{Berger}}, \bibinfo {author}
  {\bibfnamefont{P.}~\bibnamefont{Bugnon}}, \bibinfo {author}
  {\bibfnamefont{Y.~J.}\ \bibnamefont{Chang}}, \bibinfo {author}
  {\bibfnamefont{K.}~\bibnamefont{Kern}}, \bibinfo {author}
  {\bibfnamefont{A.}~\bibnamefont{Bostwick}}, \bibinfo {author}
  {\bibfnamefont{E.}~\bibnamefont{Rotenberg}}, \bibinfo {author}
  {\bibfnamefont{O.~V.}\ \bibnamefont{Yazyev}},\ and\ \bibinfo {author}
  {\bibfnamefont{M.}~\bibnamefont{Grioni}},\ }%
  \bibfield{journal}{%
  \Doi{10.1103/PhysRevLett.109.096803}{\bibinfo {journal} {Phys. Rev. Lett.}}\
  }%
  \textbf{\bibinfo {volume} {109}},\ \bibinfo {pages} {096803} (\bibinfo {year}
  {2012})%
  \bibAnnoteFile{NoStop}{Crepaldi2012}%
\bibitem{Eremeev2012}%
  \BibitemOpen
  \bibfield{author}{%
  \bibinfo {author} {\bibfnamefont{S.~V.}\ \bibnamefont{Eremeev}}, \bibinfo
  {author} {\bibfnamefont{I.~A.}\ \bibnamefont{Nechaev}}, \bibinfo {author}
  {\bibfnamefont{Y.~M.}\ \bibnamefont{Koroteev}}, \bibinfo {author}
  {\bibfnamefont{P.~M.}\ \bibnamefont{Echenique}},\ and\ \bibinfo {author}
  {\bibfnamefont{E.~V.}\ \bibnamefont{Chulkov}},\ }%
  \bibfield{journal}{%
  \Doi{10.1103/PhysRevLett.108.246802}{\bibinfo {journal} {Phys. Rev. Lett.}}\
  }%
  \textbf{\bibinfo {volume} {108}},\ \bibinfo {pages} {246802} (\bibinfo {year}
  {2012})%
  \bibAnnoteFile{NoStop}{Eremeev2012}%
\bibitem{Landolt2013}%
  \BibitemOpen
  \bibfield{author}{%
  \bibinfo {author} {\bibfnamefont{G.}~\bibnamefont{Landolt}}, \bibinfo
  {author} {\bibfnamefont{S.~V.}\ \bibnamefont{Eremeev}}, \bibinfo {author}
  {\bibfnamefont{O.~E.}\ \bibnamefont{Tereshchenko}}, \bibinfo {author}
  {\bibfnamefont{S.}~\bibnamefont{Muff}}, \bibinfo {author}
  {\bibfnamefont{B.}~\bibnamefont{Slomski}}, \bibinfo {author}
  {\bibfnamefont{K.~A.}\ \bibnamefont{Kokh}}, \bibinfo {author}
  {\bibfnamefont{M.}~\bibnamefont{Kobayashi}}, \bibinfo {author}
  {\bibfnamefont{T.}~\bibnamefont{Schmitt}}, \bibinfo {author}
  {\bibfnamefont{V.~N.}\ \bibnamefont{Strocov}}, \bibinfo {author}
  {\bibfnamefont{J.}~\bibnamefont{Osterwalder}}, \bibinfo {author}
  {\bibfnamefont{E.~V.}\ \bibnamefont{Chulkov}},\ and\ \bibinfo {author}
  {\bibfnamefont{J.~H.}\ \bibnamefont{Dil}},\ }%
  \bibfield{journal}{%
  \bibinfo {journal} {New J. Phys.}\ }%
  \textbf{\bibinfo {volume} {15}},\ \bibinfo {pages} {085022} (\bibinfo {year}
  {2013})%
  \bibAnnoteFile{NoStop}{Landolt2013}%
\bibitem{Eremeev2013}%
  \BibitemOpen
  \bibfield{author}{%
  \bibinfo {author} {\bibfnamefont{S.~V.}\ \bibnamefont{Eremeev}}, \bibinfo
  {author} {\bibfnamefont{I.~P.}\ \bibnamefont{Rusinov}}, \bibinfo {author}
  {\bibfnamefont{I.~A.}\ \bibnamefont{Nechaev}},\ and\ \bibinfo {author}
  {\bibfnamefont{E.~V.}\ \bibnamefont{Chulkov}},\ }%
  \bibfield{journal}{%
  \bibinfo {journal} {New J. Phys.}\ }%
  \textbf{\bibinfo {volume} {15}},\ \bibinfo {pages} {075015} (\bibinfo {year}
  {2013})%
  \bibAnnoteFile{NoStop}{Eremeev2013}%
\bibitem{Petersen2000}%
  \BibitemOpen
  \bibfield{author}{%
  \bibinfo {author} {\bibfnamefont{L.}~\bibnamefont{Petersen}}\ and\ \bibinfo
  {author} {\bibfnamefont{P.}~\bibnamefont{Hedeg{\aa}rd}},\ }%
  \bibfield{journal}{%
  \Doi{http://dx.doi.org/10.1016/S0039-6028(00)00441-6}{\bibinfo {journal}
  {Surf. Sci.}}\ }%
  \textbf{\bibinfo {volume} {459}},\ \bibinfo {pages} {49 } (\bibinfo {year}
  {2000})%
  \bibAnnoteFile{NoStop}{Petersen2000}%
\bibitem{Henk2003}%
  \BibitemOpen
  \bibfield{author}{%
  \bibinfo {author} {\bibfnamefont{J.}~\bibnamefont{Henk}}, \bibinfo {author}
  {\bibfnamefont{A.}~\bibnamefont{Ernst}},\ and\ \bibinfo {author}
  {\bibfnamefont{P.}~\bibnamefont{Bruno}},\ }%
  \bibfield{journal}{%
  \Doi{10.1103/PhysRevB.68.165416}{\bibinfo {journal} {Phys. Rev. B}}\ }%
  \textbf{\bibinfo {volume} {68}},\ \bibinfo {pages} {165416} (\bibinfo {year}
  {2003})%
  \bibAnnoteFile{NoStop}{Henk2003}%
\bibitem{Hoesch2004}%
  \BibitemOpen
  \bibfield{author}{%
  \bibinfo {author} {\bibfnamefont{M.}~\bibnamefont{Hoesch}}, \bibinfo {author}
  {\bibfnamefont{M.}~\bibnamefont{Muntwiler}}, \bibinfo {author}
  {\bibfnamefont{V.~N.}\ \bibnamefont{Petrov}}, \bibinfo {author}
  {\bibfnamefont{M.}~\bibnamefont{Hengsberger}}, \bibinfo {author}
  {\bibfnamefont{L.}~\bibnamefont{Patthey}}, \bibinfo {author}
  {\bibfnamefont{M.}~\bibnamefont{Shi}}, \bibinfo {author}
  {\bibfnamefont{M.}~\bibnamefont{Falub}}, \bibinfo {author}
  {\bibfnamefont{T.}~\bibnamefont{Greber}},\ and\ \bibinfo {author}
  {\bibfnamefont{J.}~\bibnamefont{Osterwalder}},\ }%
  \bibfield{journal}{%
  \Doi{10.1103/PhysRevB.69.241401}{\bibinfo {journal} {Phys. Rev. B}}\ }%
  \textbf{\bibinfo {volume} {69}},\ \bibinfo {pages} {241401} (\bibinfo {year}
  {2004})%
  \bibAnnoteFile{NoStop}{Hoesch2004}%
\bibitem{Bihlmayer2006}%
  \BibitemOpen
  \bibfield{author}{%
  \bibinfo {author} {\bibfnamefont{G.}~\bibnamefont{Bihlmayer}}, \bibinfo
  {author} {\bibfnamefont{Y.~M.}\ \bibnamefont{Koroteev}}, \bibinfo {author}
  {\bibfnamefont{P.~M.}\ \bibnamefont{Echenique}}, \bibinfo {author}
  {\bibfnamefont{E.~V.}\ \bibnamefont{Chulkov}},\ and\ \bibinfo {author}
  {\bibfnamefont{S.}~\bibnamefont{Bl\"ugel}},\ }%
  \bibfield{journal}{%
  \Doi{http://dx.doi.org/10.1016/j.susc.2006.01.098}{\bibinfo {journal} {Surf.
  Sci.}}\ }%
  \textbf{\bibinfo {volume} {600}},\ \bibinfo {pages} {3888} (\bibinfo {year}
  {2006})%
  \bibAnnoteFile{NoStop}{Bihlmayer2006}%
\bibitem{Nagano2009}%
  \BibitemOpen
  \bibfield{author}{%
  \bibinfo {author} {\bibfnamefont{M.}~\bibnamefont{Nagano}}, \bibinfo {author}
  {\bibfnamefont{A.}~\bibnamefont{Kodama}}, \bibinfo {author}
  {\bibfnamefont{T.}~\bibnamefont{Shishidou}},\ and\ \bibinfo {author}
  {\bibfnamefont{T.}~\bibnamefont{Oguchi}},\ }%
  \bibfield{journal}{%
  \bibinfo {journal} {J. Phys.: Condens. Matter}\ }%
  \textbf{\bibinfo {volume} {21}},\ \bibinfo {pages} {064239} (\bibinfo {year}
  {2009})%
  \bibAnnoteFile{NoStop}{Nagano2009}%
\bibitem{Bentmann2011}%
  \BibitemOpen
  \bibfield{author}{%
  \bibinfo {author} {\bibfnamefont{H.}~\bibnamefont{Bentmann}}, \bibinfo
  {author} {\bibfnamefont{T.}~\bibnamefont{Kuzumaki}}, \bibinfo {author}
  {\bibfnamefont{G.}~\bibnamefont{Bihlmayer}}, \bibinfo {author}
  {\bibfnamefont{S.}~\bibnamefont{Bl\"ugel}}, \bibinfo {author}
  {\bibfnamefont{E.~V.}\ \bibnamefont{Chulkov}}, \bibinfo {author}
  {\bibfnamefont{F.}~\bibnamefont{Reinert}},\ and\ \bibinfo {author}
  {\bibfnamefont{K.}~\bibnamefont{Sakamoto}},\ }%
  \bibfield{journal}{%
  \Doi{10.1103/PhysRevB.84.115426}{\bibinfo {journal} {Phys. Rev. B}}\ }%
  \textbf{\bibinfo {volume} {84}},\ \bibinfo {pages} {115426} (\bibinfo {year}
  {2011})%
  \bibAnnoteFile{NoStop}{Bentmann2011}%
\bibitem{Lee2012}%
  \BibitemOpen
  \bibfield{author}{%
  \bibinfo {author} {\bibfnamefont{H.}~\bibnamefont{Lee}}\ and\ \bibinfo
  {author} {\bibfnamefont{H.~J.}\ \bibnamefont{Choi}},\ }%
  \bibfield{journal}{%
  \Doi{10.1103/PhysRevB.86.045437}{\bibinfo {journal} {Phys. Rev. B}}\ }%
  \textbf{\bibinfo {volume} {86}},\ \bibinfo {pages} {045437} (\bibinfo {year}
  {2012})%
  \bibAnnoteFile{NoStop}{Lee2012}%
\bibitem{Hortamani2012}%
  \BibitemOpen
  \bibfield{author}{%
  \bibinfo {author} {\bibfnamefont{M.}~\bibnamefont{Hortamani}}\ and\ \bibinfo
  {author} {\bibfnamefont{R.}~\bibnamefont{Wiesendanger}},\ }%
  \bibfield{journal}{%
  \Doi{10.1103/PhysRevB.86.235437}{\bibinfo {journal} {Phys. Rev. B}}\ }%
  \textbf{\bibinfo {volume} {86}},\ \bibinfo {pages} {235437} (\bibinfo {year}
  {2012})%
  \bibAnnoteFile{NoStop}{Hortamani2012}%
\bibitem{Premper2007}%
  \BibitemOpen
  \bibfield{author}{%
  \bibinfo {author} {\bibfnamefont{J.}~\bibnamefont{Premper}}, \bibinfo
  {author} {\bibfnamefont{M.}~\bibnamefont{Trautmann}}, \bibinfo {author}
  {\bibfnamefont{J.}~\bibnamefont{Henk}},\ and\ \bibinfo {author}
  {\bibfnamefont{P.}~\bibnamefont{Bruno}},\ }%
  \bibfield{journal}{%
  \Doi{10.1103/PhysRevB.76.073310}{\bibinfo {journal} {Phys. Rev. B}}\ }%
  \textbf{\bibinfo {volume} {76}},\ \bibinfo {pages} {073310} (\bibinfo {year}
  {2007})%
  \bibAnnoteFile{NoStop}{Premper2007}%
\bibitem{Moreschini2009}%
  \BibitemOpen
  \bibfield{author}{%
  \bibinfo {author} {\bibfnamefont{L.}~\bibnamefont{Moreschini}}, \bibinfo
  {author} {\bibfnamefont{A.}~\bibnamefont{Bendounan}}, \bibinfo {author}
  {\bibfnamefont{H.}~\bibnamefont{Bentmann}}, \bibinfo {author}
  {\bibfnamefont{M.}~\bibnamefont{Assig}}, \bibinfo {author}
  {\bibfnamefont{K.}~\bibnamefont{Kern}}, \bibinfo {author}
  {\bibfnamefont{F.}~\bibnamefont{Reinert}}, \bibinfo {author}
  {\bibfnamefont{J.}~\bibnamefont{Henk}}, \bibinfo {author}
  {\bibfnamefont{C.~R.}\ \bibnamefont{Ast}},\ and\ \bibinfo {author}
  {\bibfnamefont{M.}~\bibnamefont{Grioni}},\ }%
  \bibfield{journal}{%
  \Doi{10.1103/PhysRevB.80.035438}{\bibinfo {journal} {Phys. Rev. B}}\ }%
  \textbf{\bibinfo {volume} {80}},\ \bibinfo {pages} {035438} (\bibinfo {year}
  {2009})%
  \bibAnnoteFile{NoStop}{Moreschini2009}%
\bibitem{Krasovskii2011}%
  \BibitemOpen
  \bibfield{author}{%
  \bibinfo {author} {\bibfnamefont{E.~E.}\ \bibnamefont{Krasovskii}}\ and\
  \bibinfo {author} {\bibfnamefont{E.~V.}\ \bibnamefont{Chulkov}},\ }%
  \bibfield{journal}{%
  \Doi{10.1103/PhysRevB.83.155401}{\bibinfo {journal} {Phys. Rev. B}}\ }%
  \textbf{\bibinfo {volume} {83}},\ \bibinfo {pages} {155401} (\bibinfo {year}
  {2011})%
  \bibAnnoteFile{NoStop}{Krasovskii2011}%
\bibitem{Tamm1932}%
  \BibitemOpen
  \bibfield{author}{%
  \bibinfo {author} {\bibfnamefont{I.}~\bibnamefont{Tamm}},\ }%
  \bibfield{journal}{%
  \bibinfo {journal} {Phys. Z. Soviet Union}\ }%
  \textbf{\bibinfo {volume} {1}},\ \bibinfo {pages} {733} (\bibinfo {year}
  {1932})%
  \bibAnnoteFile{NoStop}{Tamm1932}%
\bibitem{Forstmann1970}%
  \BibitemOpen
  \bibfield{author}{%
  \bibinfo {author} {\bibfnamefont{F.}~\bibnamefont{Forstmann}},\ }%
  \bibfield{journal}{%
  \bibinfo {journal} {Z. Phys.}\ }%
  \textbf{\bibinfo {volume} {235}},\ \bibinfo {pages} {69} (\bibinfo {year}
  {1970})%
  \bibAnnoteFile{NoStop}{Forstmann1970}%
\bibitem{Krasovskii1997}%
  \BibitemOpen
  \bibfield{author}{%
  \bibinfo {author} {\bibfnamefont{E.~E.}\ \bibnamefont{Krasovskii}}\ and\
  \bibinfo {author} {\bibfnamefont{W.}~\bibnamefont{Schattke}},\ }%
  \bibfield{journal}{%
  \Doi{10.1103/PhysRevB.56.12874}{\bibinfo {journal} {Phys. Rev. B}}\ }%
  \textbf{\bibinfo {volume} {56}},\ \bibinfo {pages} {12874} (\bibinfo {year}
  {1997})%
  \bibAnnoteFile{NoStop}{Krasovskii1997}%
\bibitem{Winkler2003}%
  \BibitemOpen
  \bibfield{author}{%
  \bibinfo {author} {\bibfnamefont{R.}~\bibnamefont{Winkler}},\ }%
  \emph{\bibinfo {title} {Spin–orbit coupling effects in two-dimensional
  electron and hole systems}}\ (\bibinfo {publisher} {Springer},\ \bibinfo
  {year} {2003})%
  \bibAnnoteFile{NoStop}{Winkler2003}%
\bibitem{Oguchi2009}%
  \BibitemOpen
  \bibfield{author}{%
  \bibinfo {author} {\bibfnamefont{T.}~\bibnamefont{Oguchi}}\ and\ \bibinfo
  {author} {\bibfnamefont{T.}~\bibnamefont{Shishidou}},\ }%
  \bibfield{journal}{%
  \bibinfo {journal} {J. Phys.: Condens. Matter}\ }%
  \textbf{\bibinfo {volume} {21}},\ \bibinfo {pages} {092001} (\bibinfo {year}
  {2009})%
  \bibAnnoteFile{NoStop}{Oguchi2009}%
\bibitem{KSS99}%
  \BibitemOpen
  \bibfield{author}{%
  \bibinfo {author} {\bibfnamefont{E.~E.}\ \bibnamefont{Krasovskii}}, \bibinfo
  {author} {\bibfnamefont{F.}~\bibnamefont{Starrost}},\ and\ \bibinfo {author}
  {\bibfnamefont{W.}~\bibnamefont{Schattke}},\ }%
  \bibfield{journal}{%
  \Doi{10.1103/PhysRevB.59.10504}{\bibinfo {journal} {Phys. Rev. B}}\ }%
  \textbf{\bibinfo {volume} {59}},\ \bibinfo {pages} {10504} (\bibinfo {year}
  {1999})%
  \bibAnnoteFile{NoStop}{KSS99}%
\bibitem{Lobo-Checa2011}%
  \BibitemOpen
  \bibfield{author}{%
  \bibinfo {author} {\bibfnamefont{J.}~\bibnamefont{Lobo-Checa}}, \bibinfo
  {author} {\bibfnamefont{J.~E.}\ \bibnamefont{Ortega}}, \bibinfo {author}
  {\bibfnamefont{A.}~\bibnamefont{Mascaraque}}, \bibinfo {author}
  {\bibfnamefont{E.~G.}\ \bibnamefont{Michel}},\ and\ \bibinfo {author}
  {\bibfnamefont{E.~E.}\ \bibnamefont{Krasovskii}},\ }%
  \bibfield{journal}{%
  \Doi{10.1103/PhysRevB.84.245419}{\bibinfo {journal} {Phys. Rev. B}}\ }%
  \textbf{\bibinfo {volume} {84}},\ \bibinfo {pages} {245419} (\bibinfo {year}
  {2011})%
  \bibAnnoteFile{NoStop}{Lobo-Checa2011}%
\bibitem{Borghetti2012}%
  \BibitemOpen
  \bibfield{author}{%
  \bibinfo {author} {\bibfnamefont{P.}~\bibnamefont{Borghetti}}, \bibinfo
  {author} {\bibfnamefont{J.}~\bibnamefont{Lobo-Checa}}, \bibinfo {author}
  {\bibfnamefont{E.}~\bibnamefont{Goiri}}, \bibinfo {author}
  {\bibfnamefont{A.}~\bibnamefont{Mugarza}}, \bibinfo {author}
  {\bibfnamefont{F.}~\bibnamefont{Schiller}}, \bibinfo {author}
  {\bibfnamefont{J.~E.}\ \bibnamefont{Ortega}},\ and\ \bibinfo {author}
  {\bibfnamefont{E.~E.}\ \bibnamefont{Krasovskii}},\ }%
  \bibfield{journal}{%
  \bibinfo {journal} {J. Phys.: Condens. Matter}\ }%
  \textbf{\bibinfo {volume} {24}},\ \bibinfo {pages} {395006} (\bibinfo {year}
  {2012})%
  \bibAnnoteFile{NoStop}{Borghetti2012}%
\bibitem{Note1}%
  \BibitemOpen
  \bibinfo {note} {For the crystal structure of the Ag$_2$Bi surface alloy and
  the properties of the surface state see, e.g., Refs.~\onlinecite
  {Meier2008,Gierz2010,Ast2007,Ast2008}}%
  \bibAnnoteFile{NoStop}{Note1}%
\bibitem{Kim2011}%
  \BibitemOpen
  \bibfield{author}{%
  \bibinfo {author} {\bibfnamefont{S.}~\bibnamefont{Kim}}, \bibinfo {author}
  {\bibfnamefont{M.}~\bibnamefont{Ye}}, \bibinfo {author}
  {\bibfnamefont{K.}~\bibnamefont{Kuroda}}, \bibinfo {author}
  {\bibfnamefont{Y.}~\bibnamefont{Yamada}}, \bibinfo {author}
  {\bibfnamefont{E.~E.}\ \bibnamefont{Krasovskii}}, \bibinfo {author}
  {\bibfnamefont{E.~V.}\ \bibnamefont{Chulkov}}, \bibinfo {author}
  {\bibfnamefont{K.}~\bibnamefont{Miyamoto}}, \bibinfo {author}
  {\bibfnamefont{M.}~\bibnamefont{Nakatake}}, \bibinfo {author}
  {\bibfnamefont{T.}~\bibnamefont{Okuda}}, \bibinfo {author}
  {\bibfnamefont{Y.}~\bibnamefont{Ueda}}, \bibinfo {author}
  {\bibfnamefont{K.}~\bibnamefont{Shimada}}, \bibinfo {author}
  {\bibfnamefont{H.}~\bibnamefont{Namatame}}, \bibinfo {author}
  {\bibfnamefont{M.}~\bibnamefont{Taniguchi}},\ and\ \bibinfo {author}
  {\bibfnamefont{A.}~\bibnamefont{Kimura}},\ }%
  \bibfield{journal}{%
  \Doi{10.1103/PhysRevLett.107.056803}{\bibinfo {journal} {Phys. Rev. Lett.}}\
  }%
  \textbf{\bibinfo {volume} {107}},\ \bibinfo {pages} {056803} (\bibinfo {year}
  {2011})%
  \bibAnnoteFile{NoStop}{Kim2011}%
\bibitem{KOH77}%
  \BibitemOpen
  \bibfield{author}{%
  \bibinfo {author} {\bibfnamefont{D.~D.}\ \bibnamefont{Koelling}}\ and\
  \bibinfo {author} {\bibfnamefont{B.~N.}\ \bibnamefont{Harmon}},\ }%
  \bibfield{journal}{%
  \bibinfo {journal} {J. Phys. C}\ }%
  \textbf{\bibinfo {volume} {10}},\ \bibinfo {pages} {3107} (\bibinfo {year}
  {1977})%
  \bibAnnoteFile{NoStop}{KOH77}%
\bibitem{MCD80}%
  \BibitemOpen
  \bibfield{author}{%
  \bibinfo {author} {\bibfnamefont{A.~H.}\ \bibnamefont{MacDonald}}, \bibinfo
  {author} {\bibfnamefont{W.~E.}\ \bibnamefont{Picket}},\ and\ \bibinfo
  {author} {\bibfnamefont{D.~D.}\ \bibnamefont{Koelling}},\ }%
  \bibfield{journal}{%
  \bibinfo {journal} {J. Phys. C}\ }%
  \textbf{\bibinfo {volume} {13}},\ \bibinfo {pages} {2675} (\bibinfo {year}
  {1980})%
  \bibAnnoteFile{NoStop}{MCD80}%
\bibitem{Krasovskii1993}%
  \BibitemOpen
  \bibfield{author}{%
  \bibinfo {author} {\bibfnamefont{E.~E.}\ \bibnamefont{Krasovskii}}, \bibinfo
  {author} {\bibfnamefont{V.~V.}\ \bibnamefont{Nemoshkalenko}},\ and\ \bibinfo
  {author} {\bibfnamefont{V.~N.}\ \bibnamefont{Antonov}},\ }%
  \bibfield{journal}{%
  \bibinfo {journal} {Z. Phys. B}\ }%
  \textbf{\bibinfo {volume} {91}} (\bibinfo {year} {1993})%
  \bibAnnoteFile{NoStop}{Krasovskii1993}%
\bibitem{Azpiroz2013}%
  \BibitemOpen
  \bibfield{author}{%
  \bibinfo {author} {\bibfnamefont{J.}~\bibnamefont{Iba\~nez Azpiroz}},
  \bibinfo {author} {\bibfnamefont{A.}~\bibnamefont{Bergara}}, \bibinfo
  {author} {\bibfnamefont{E.~Y.}\ \bibnamefont{Sherman}},\ and\ \bibinfo
  {author} {\bibfnamefont{A.}~\bibnamefont{Eiguren}},\ }%
  \bibfield{journal}{%
  \Doi{10.1103/PhysRevB.88.125404}{\bibinfo {journal} {Phys. Rev. B}}\ }%
  \textbf{\bibinfo {volume} {88}},\ \bibinfo {pages} {125404} (\bibinfo {year}
  {2013})%
  \bibAnnoteFile{NoStop}{Azpiroz2013}%
\bibitem{Nechaev2010}%
  \BibitemOpen
  \bibfield{author}{%
  \bibinfo {author} {\bibfnamefont{I.~A.}\ \bibnamefont{Nechaev}}, \bibinfo
  {author} {\bibfnamefont{P.~M.}\ \bibnamefont{Echenique}},\ and\ \bibinfo
  {author} {\bibfnamefont{E.~V.}\ \bibnamefont{Chulkov}},\ }%
  \bibfield{journal}{%
  \Doi{10.1103/PhysRevB.81.195112}{\bibinfo {journal} {Phys. Rev. B}}\ }%
  \textbf{\bibinfo {volume} {81}},\ \bibinfo {pages} {195112} (\bibinfo {year}
  {2010})%
  \bibAnnoteFile{NoStop}{Nechaev2010}%
\end{thebibliography}
\end{document}